\documentclass[aps, prd, amsmath, floats, floatfix, twocolumn, nofootinbib, superscriptaddress, showpacs]{revtex4-1}
\usepackage{graphicx}
\usepackage{color}
\usepackage{latexsym}

\newcommand{\be}{\begin{eqnarray}}
\newcommand{\ee}{\end{eqnarray}}

\begin{document}

\title{Testing the no-hair theorem with the continuum-fitting\\and the iron line methods: a short review}

\author{Cosimo Bambi}
\affiliation{Center for Field Theory and Particle Physics and Department of Physics, Fudan University, 200433 Shanghai, China}
\affiliation{Theoretical Astrophysics, Eberhard-Karls Universit\"at T\"ubingen, 72076 T\"ubingen, Germany}

\author{Jiachen Jiang}
\affiliation{Center for Field Theory and Particle Physics and Department of Physics, Fudan University, 200433 Shanghai, China}

\author{James F. Steiner}
\altaffiliation{Einstein Fellow}
\affiliation{MIT Kavli Institute, Cambridge, MA 02139, United States}

\date{\today}

\begin{abstract}
The continuum-fitting and the iron line methods are leading techniques capable of probing the spacetime geometry around astrophysical black hole candidates and testing the no-hair theorem. In the present paper, we review the two approaches, from the astrophysical models and their assumptions, to the constraining power with present and future facilities.
\end{abstract}

\maketitle


\section{Introduction}

In 4-dimensional general relativity, black holes are described by the Kerr-Newman solution and are completely characterized by only three parameters, associated with the mass $M$, the spin angular momentum $J$, and the electric charge $Q$ of the compact object. 
The ``no-hair theorem'' describes that black holes have only these three asymptotic charges and no more.  This theorem was pioneered by Israel~\cite{h1}, Carter~\cite{h2}, and Robinson~\cite{h3} almost 50~years ago, and its final version is still a work in progress (for a review, see e.g. Ref.~\cite{h4}). The name no-hair is to indicate that black holes have no features (hairs), although to be precise, black holes can only have three hairs ($M$, $J$, and $Q$). 
The fact that there is only the Kerr-Newman solution is the result of the ``uniqueness theorem''. In the context of tests of the Kerr metric and of general relativity, both theorems are relevant.  E.g., as a matter of principle, one may have different classes of black holes (each with characteristic $M$, $J$, and $Q$ hairs), thus violating the uniqueness theorem, without any violation of the no-hair theorem.

The no-hair theorem holds under specific assumptions. For instance, it is violated in more than 4 dimensions~\cite{h5}. ``Hairy'' black holes naturally arise in the presence of non-Abelian gauge fields~\cite{h6,h7}. In some alternative theories of gravity, the theorem may still holds, and an example is a simple scalar-tensor theory, in which the black hole solutions are the same as in general relativity~\cite{v2-sf12}. In other frameworks, the black hole solutions of general relativity may still be solutions of the new field equations, but  their uniqueness is not guaranteed. A relevant example of violation of the no-hair theorem is  presented in the recent paper by Herdeiro and Radu~\cite{v2-hr14}.

In general, we can distinguish two kinds of hairs, called, respectively, primary and secondary hairs. {\it Primary hairs} are real hairs of the black hole: if such hairs were to exist, then $M$, $J$, and $Q$ would not completely characterize the compact object, and one or more additional parameters would be necessary. An example is a 5-dimensional Myers-Perry black hole: it is the 5-dimensional generalization of Kerr black holes and it has two angular momenta, so one more hair: $J'$~\cite{h5}. {\it Secondary hairs} are instead related to some new charge that is common to all black holes. For instance, in Einstein-dilaton-Gauss-Bonnet gravity, a black hole has a scalar charge proportional to the volume integral of the Gauss-Bonnet invariant~\cite{v2-kmrtw95,v2-ms93,v2-pc09,h8,v2-mpgf15}; that is, the scalar charge is determined by the black hole mass and it is not an additional degree of freedom.

Astrophysical black hole candidates are dark and compact objects that can be naturally interpreted as the black holes of general relativity or instead they could be something else only in the presence of new physics~\cite{ramesh}. Stellar-mass black hole candidates in X-ray binaries are compact objects that are too massive to be neutron stars for any plausible matter equation of state~\cite{bh2}. Their mass can be measured by dynamical methods, by studying the orbital motion of the companion star, usually with optical observations~\cite{bh3}. The size of these objects follows from the short time scale variability. Supermassive black hole candidates in galactic nuclei are too massive, too compact, and too old to be cluster of neutron stars.  The expected lifetime due to evaporation or physical collision would be shorter than the age of these systems~\cite{bh4}.

It is worth noting that the spacetime geometry around astrophysical black holes formed by gravitational collapse should be well described by the stationary Kerr solution with $|J/M^2| \le 1$. In general relativity, initial deviations from the Kerr metric are indeed quickly radiated away through the emission of gravitational waves (``black holes rapidly go bald'')~\cite{k1}. The equilibrium electric charge is extremely small for macroscopic objects and irrelevant in the spacetime geometry, so we can safely assume that $Q=0$~\cite{v2-g75,v2-gj69,v2-bz77,k2}. The presence of an accretion disk is normally a completely negligible influence, because the disk mass is several orders of magnitude smaller than the mass of the black hole~\cite{v2-bcp14,k3}. The spin is constrained to be $|J/M^2| \le 1$ because, otherwise, there would be no black hole but a naked singularity. While one may argue that new physics may eliminate the singularity~\cite{2-1}, and that spin measurements should not assume any {\it a priori} constraint~\cite{2-2a,2-2b,2-2c,2-2d}, it appears today that Kerr ``superspinars'' are not viable black hole candidates. First, it does not appear posssible to create a Kerr naked singularity~\cite{2-3a,2-3b,2-3c}. Second, and more importantly, even if created, a similar object would be very unstable and would quickly decay~\cite{2-4}.

Testing the no-hair theorem with astrophysical black hole candidates is often interpreted to mean testing the Kerr metric.  However, strictly speaking, a non-Kerr black hole might still be completely characterized by $M$ and $J$ only. For instance, this would be the case for uncharged black holes with secondary hairs. The aim of this paper is to briefly review present attempts to test the no-hair theorem with the continuum-fitting and the iron line methods. More details on tests of the Kerr metric with these and other electromagnetic techniques can be found in~\cite{review-s,review-l} and references therein. Tests with gravitational waves are reviewed in~\cite{v2-ys13,v2-b15}. The continuum-fitting and the iron line methods are today the primary tools for probing the spacetime geometry around astrophysical black hole candidates. They are presently being used to measure the spin under the assumption of the Kerr metric, and by extension they can be used to test the no-hair theorem.

Continuum-fitting refers to the analysis of the thermal spectrum of thin disks. These measurements are relatively robust, because the underlying physics is well known. However, the spectrum of thin disks has a simple shape and it is accordingly quite challenging to constrain deviations from the Kerr solution. Still, some interesting bounds can be obtained from the observations of black hole candidates that look like fast-rotating Kerr black holes.

The iron-line method involves analysis of the X-ray reflection emission component. Such emission results from the illumination of the disk by a hot corona and consists of a broadband continuum plus several fluorescent lines.  Most pronounced in the observational band is the iron line; the detailed modeling of the profile of this line (and similar features) readily provides information on the spacetime geometry. This technique is potentially more powerful than the continuum-fitting approach, because the profile of the line has distinctive structure. However, there are systematic uncertainties in the methodology which are not understood at more than a cursory level.  In particular, the exact geometry of the corona is not known. The continuum-fitting method has enormous statistical precision and is generally limited by external knowledge of the system mass and distance.  By contrast, the iron line method is generally limited by signal in the line, as well as systematic limitations of the model. In the future, with X-ray observatories with larger effective areas, the iron line analysis will be able to do a better job if the underpinning model and system geometry are known to high precision.

In the next sections, we will use natural units in which $G_{\rm N} = c = 1$, unless stated otherwise, and adopt a metric signature $(-+++)$.

\section{Test-metrics}

In tests of the no-hair theorem with black hole candidates, it is common to use a model-independent approach that closely resembles the PPN (Parametrized Post-Newtonian) formalism. In the PPN case, one wants to test the Schwarzschild solution in the weak field limit~\cite{ppn68}. The starting point is to assume the most general static, spherically symmetric, and asymptotically flat line element in the weak field limit $(M/r \ll 1)$. In isotropic coordinates, it reads
\be\label{eq-ppn}
ds^2 &=& - \left(1 - \frac{2 M}{r} + \beta\frac{2 M^2}{r^2} + . . .  \right) dt^2 \nonumber\\
&& + \left(1 + \gamma \frac{2 M}{r} + . . .  \right) \left(dx^2 + dy^2 + dz^2 \right) \, ,
\ee
where $\beta$ and $\gamma$ are free parameters to be measured by experiments. The only spherically symmetric vacuum solution of Einstein's field equations is the Schwarzschild metric and it requires $\beta = \gamma = 1$. Current observational data in the Solar System constrain $\beta$ and $\gamma$ to be 1 with a precision, respectively, of $10^{-4}$ and $10^{-5}$, confirming the Schwarzschild solution at this level of accuracy~\cite{ppn03,ppn04}.

With the same spirit, one can write a more general metric than the Kerr solution and that includes the Kerr solution as a special case. In addition to $M$ and $J$, the line element will have a number of {\it deformation parameters} that, like $\beta$ and $\gamma$ in the PPN metric, parametrize possible deviations from the predictions of general relativity. These deformation parameters must be measured by observations and, {\it a posterior}, one can check whether they vanish, as required by the Kerr black hole hypothesis. 
In principle, one should use a general metric capable of mimicking black hole solutions in alternative theories for a proper choice of the value of its deformation parameters. In practice, this is quite challenging, and so there are several efforts to find a successful parametrization that can reproduce the known non-Kerr black hole solutions using the smallest possible number of extra parameters.

A very popular choice is the Johannsen-Psaltis metric~\cite{jp-m}. It is not a solution of a specific alternative theory of gravity. In Boyer-Lindquist coordinates, the line element reads
\begin{widetext}
\be\label{eq-jp}
ds^2 &=& - \left(1 - \frac{2 M r}{\Sigma}\right)\left(1 + h\right) dt^2
 - \frac{4 M a r \sin^2\theta}{\Sigma}\left(1 + h\right) dt d\phi 
+ \frac{\Sigma \left(1 + h\right)}{\Delta + h a^2 \sin^2\theta} dr^2
+ \Sigma d\theta^2 \nonumber\\
&& + \left[r^2 + a^2 + \frac{2 a^2 M r \sin^2\theta}{\Sigma}
+ \frac{a^2 \left(\Sigma + 2 M r\right) \sin^2\theta}{\Sigma} h \right] 
\sin^2\theta d\phi^2 \, ,
\ee
\end{widetext}
where $a=J/M$ is the specific spin, $\Sigma = r^2 + a^2 \cos^2\theta$, $\Delta = r^2 - 2 M r + a^2$, and 
\be
h &=& \sum_{k=0}^{+\infty} \left(\epsilon_{2k}
+ \epsilon_{2k+1} \frac{M r}{\Sigma}\right)\left(\frac{M^2}{\Sigma}
\right)^k \, .
\ee
The metric has an infinite set of deformation parameters $\{\epsilon_k\}$ and it reduces to the Kerr solution when all the deformation parameters vanish. However, $\epsilon_0$ must vanish to recover the correct Newtonian limit, while $\epsilon_1$ and $\epsilon_2$ are already strongly constrained by experiments in the Solar System. The simplest non-trivial metric is thus that with $\epsilon_3$ free and with all the other deformation parameters set to zero.

An extension of the Johannsen-Psaltis metric is the Cardoso-Pani-Rico parametrization~\cite{cpr-m}. Its line element is
\begin{widetext}
\be\label{eq-cpr}
ds^2 &=& - \left(1 - \frac{2 M r}{\Sigma}\right)\left(1 + h^t\right) dt^2
- 2 a \sin^2\theta \left[\sqrt{\left(1 + h^t\right)\left(1 + h^r\right)} 
- \left(1 - \frac{2 M r}{\Sigma}\right)\left(1 + h^t\right)\right] dt d\phi \nonumber\\
&& + \frac{\Sigma \left(1 + h^r\right)}{\Delta + h^r a^2 \sin^2\theta} dr^2
+ \Sigma d\theta^2 
+ \sin^2\theta \left\{\Sigma + a^2 \sin^2\theta \left[ 2 \sqrt{\left(1 + h^t\right)
\left(1 + h^r\right)} - \left(1 - \frac{2 M r}{\Sigma}\right)
\left(1 + h^t\right)\right]\right\} d\phi^2 \, , \quad
\ee
\end{widetext}
where 
\be
h^t &=& \sum_{k=0}^{+\infty} \left(\epsilon_{2k}^t 
+ \epsilon_{2k+1}^t \frac{M r}{\Sigma}\right)\left(\frac{M^2}{\Sigma}
\right)^k\, , \\
h^r &=& \sum_{k=0}^{+\infty} \left(\epsilon_{2k}^r
+ \epsilon_{2k+1}^r \frac{M r}{\Sigma}\right)\left(\frac{M^2}{\Sigma}
\right)^k \, .
\ee
Now there are two infinite sets of deformation parameters, $\{\epsilon_k^t\}$ and $\{\epsilon_k^r\}$. The Johannsen-Psaltis metric is recovered when $\epsilon_k^t = \epsilon_k^r$ for any $k$, and the Kerr metric when $\epsilon_k^t = \epsilon_k^r = 0$ for any $k$.

Other tests-metrics have been proposed, see e.g. Refs.~\cite{mm1,mm2,mm3,mm4,v2-mm}. For the moment, there is not a general formalism as the PPN metric, as the present parametrizations cannot account for fully arbitrary deviations from Kerr and instead represent specific deviation classes. Accordingly, any proposal has its advantages and disadvantages, and the choice of the metric may thus be determined by the particular study that one has in mind.

\section{Continuum-fitting method}

The continuum-fitting method involves analysis of the thermal spectrum of thin accretion disks. For the moment, it has been used predominantly for stellar-mass black hole candidates, because the temperature of the disk scales as $T \propto M^{-0.25}$ and the spectrum is in the X-ray band for stellar-mass black hole candidates and in the UV/optical band for supermassive black hole candidates. In the latter case, extinction and dust absorption limit the ability to make an accurate measurement.

\subsection{Novikov-Thorne model}

The Novikov-Thorne model is the standard framework to describe geometrically thin and optically thick accretion disks~\cite{ntm,ntm2}, and it is the relativistic generalization of the Shakura-Sunyaev model~\cite{ssm}. Accretion is possible because viscous magnetic/turbulent stresses and radiation transport energy and angular momentum outwards. The model requires that the spacetime is stationary, axisymmetric and asymptotically flat, and that the disk is axisymmetric and in a steady state. The model assumes that the disk is in the equatorial plane, that the particles of the disk follows nearly geodesic circular orbits, and that the radial heat transport is negligible compared to the energy radiated from the surface of the disk. The time-averaged radial structure of the disk follows from the conservation laws for the rest-mass, angular momentum, and energy~\cite{ntm2}. The radius-independent time-averaged mass accretion rate $\dot{M}$, the time-averaged energy flux $\mathcal{F}(r)$ from the surface of the disk, and the time-average torque $W^r_\phi(r)$ are given by~\cite{ntm2}
\be
\dot{M} &=& - 2 \pi \sqrt{-G} \tilde{\Sigma} u^r = {\rm const.} \, , \\\label{fluxeq}
\mathcal{F}(r) &=& \frac{\dot{M}}{4 \pi M^2} F(r) \, , \\
W^r_\phi(r) &=& \frac{\dot{M}}{2\pi M^2} 
\frac{\Omega L_z - E}{\partial_r \Omega} F(r) \, .
\ee
Here $\tilde{\Sigma}$ is the surface density, $u^r$ is the radial 4-velocity of the particles of the gas, $G = - \alpha^2 g_{rr} g_{\phi\phi}$ is the determinant of the near equatorial plane metric, where $\alpha^2 = g_{t\phi}^2/g_{\phi\phi} - g_{tt}$ is the lapse function. $E$, $L_z$, and $\Omega$ are, respectively, the specific energy, the axial component of the specific angular momentum, and the angular velocity for equatorial circular geodesics. $F(r)$ is the dimensionless function
\be
\hspace{-0.5cm}
F(r) = - \frac{\partial_r \Omega}{(E - \Omega L)^2} 
\frac{M^2}{\sqrt{-G}}
\int_{r_{\rm in}}^{r} (E - \Omega L) 
(\partial_\rho L) d\rho \, ,
\ee
where $r_{\rm in}$ is the inner radius of the accretion disk. A crucial assumption in the Novikov-Thorne model is that the inner edge of the disk is at the radius of the innermost stable circular orbit (ISCO).

The disk is in local thermal equilibrium and therefore its emission is blackbody-like and we can define an effective temperature $T_{\rm eff} (r)$ by $\mathcal{F}(r) = \sigma T^4_{\rm eff}$, where $\sigma$ is the Stefan-Boltzmann constant. Since the temperature of the disk near the inner edge can be high, up to $\sim 10^7$~K for stellar-mass black hole candidates, corrections for non-blackbody effects can be important. This is usually taken into account by introducing the color correction term (or hardening factor) $f_{\rm col}$, which is largely due to electron scattering in the disk and in practice is obtained from disk atmosphere models~\cite{davis1,davis2}. The color temperature is defined as $T_{\rm col} (r) = f_{\rm col} T_{\rm eff}$. The local specific intensity of the radiation emitted by the disk is (reintroducing the speed of light $c$)
\be\label{eq-i-bb}
I_{\rm e}(\nu_{\rm e}) = \frac{2 h \nu^3_{\rm e}}{c^2} \frac{1}{f_{\rm col}^4} 
\frac{\Upsilon}{\exp\left(\frac{h \nu_{\rm e}}{k_{\rm B} T_{\rm col}}\right) - 1} \, ,
\ee
where $\nu_{\rm e}$ is the photon frequency, $h$ is the Planck's constant, $k_{\rm B}$ is the Boltzmann constant, and $\Upsilon$ is a function of the angle between the wavevector of the photon emitted by the disk and the normal of the disk surface, say $\xi$. The two most common options are $\Upsilon = 1$ (isotropic emission) and $\Upsilon = \frac{1}{2} + \frac{3}{4} \cos\xi$ (limb-darkened emission). The choice of the form of $\Upsilon$ affects the final measurement, but the impact is small. It is one of the uncertainties in the model.

The calculation of the thermal spectrum of thin accretion disks has been extensively discussed in the literature; see e.g.~\cite{cfm2,c-cfm2} and references therein. The photon flux number density as measured by a distant observer is
\begin{widetext}
\be\label{eq-n2}
N_{E_{\rm obs}} &=&
\frac{1}{E_{\rm obs}} \int I_{\rm obs}(\nu) d \Omega_{\rm obs} = 
\frac{1}{E_{\rm obs}} \int g^3 I_{\rm e}(\nu_{\rm e}) d \Omega_{\rm obs} =  
A_1 \left(\frac{E_{\rm obs}}{\rm keV}\right)^2
\int \frac{1}{M^2} \frac{\Upsilon dXdY}{\exp\left[\frac{A_2}{g F^{1/4}} 
\left(\frac{E_{\rm obs}}{\rm keV}\right)\right] - 1} \, ,
\ee
where (reintroducing the constants $G_{\rm N}$ and $c$ and correcting a typo in Eq.~(6) in Ref.~\cite{c-cfm2}, where the Planck's constant $h$ in $A_1$ appears without $3/2$)
\be
A_1 &=&  \frac{2 \left({\rm keV}\right)^2}{f_{\rm col}^4} 
\left(\frac{G_{\rm N} M}{c^3 h^{3/2} D}\right)^2 = 
\frac{0.07205}{f_{\rm col}^4} 
\left(\frac{M}{M_\odot}\right)^2 
\left(\frac{\rm kpc}{D}\right)^2 \, 
{\rm \gamma \, keV^{-1} \, cm^{-2} \, s^{-1}} \, , \nonumber\\
A_2 &=&  \left(\frac{\rm keV}{k_{\rm B} f_{\rm col}}\right) 
\left(\frac{G_{\rm N} M}{c^3}\right)^{1/2}
\left(\frac{4 \pi \sigma}{\dot{M}}\right)^{1/4} = 
\frac{0.1331}{f_{\rm col}} 
\left(\frac{\rm 10^{18} \, g \, s^{-1}}{\dot{M}}\right)^{1/4}
\left(\frac{M}{M_\odot}\right)^{1/2} \, .
\ee
\end{widetext}
$I_{\rm obs}$, $E_{\rm obs}$, and $\nu$ are, respectively, the specific intensity of the radiation, the photon energy, and the photon frequency measured by the distant observer. $X$ and $Y$ are the coordinates of the position of the photon on the sky, $d\Omega_{\rm obs} = dX dY / D^2$ is the element of the solid angle subtended by the image of the disk on the observer's sky, and $D$ is the distance of the source. $g$ is the redshift factor
\be\label{eq-red}
g = \frac{E_{\rm obs}}{E_{\rm e}} = \frac{\nu}{\nu_{\rm e}} = 
\frac{k_\alpha u^{\alpha}_{\rm obs}}{k_\beta u^{\beta}_{\rm e}}\, ,
\ee
where $E_{\rm e} = h \nu_{\rm e}$, $k^\alpha$ is the 4-momentum of the photon, $u^{\alpha}_{\rm obs} = (1,0,0,0)$ is the 4-velocity of the distant observer, and $u^{\alpha}_{\rm e} = (u^t_{\rm e},0,0,
\Omega u^t_{\rm e})$ is the 4-velocity of the emitter. $I_{\rm e}(\nu_{\rm e})/\nu_{\rm e}^3 = I_{\rm obs} (\nu_{\rm obs})/\nu^3$ follows from Liouville's theorem.

Using the normalization condition $g_{\mu\nu}u^{\mu}_{\rm e}u^{\nu}_{\rm e} = -1$, we have
\be
u^t_{\rm e} = \frac{1}{\sqrt{-g_{tt} - 2 g_{t\phi} \Omega - g_{\phi\phi} \Omega^2}} \, ,
\ee
and therefore
\be\label{eq-red-g}
g = \frac{\sqrt{-g_{tt} - 2 g_{t\phi} \Omega - g_{\phi\phi} \Omega^2}}{1 + \lambda \Omega} \, ,
\ee
where $\lambda = k_\phi/k_t$ is a constant of the motion along the photon path. Doppler boosting and gravitational redshift are taken into account by the redshift factor $g$.

In Eq.~(\ref{eq-n2}), $g$ and $F$ depends on the emission point in the disk, $r_{\rm e}$. The integral thus requires ray tracing calculations to relate any point of emission in the disk to the point of detection in the plane of the distant observer. In other words, it is necessary to have $g = g(X,Y)$ and $F = F(X,Y)$. In the Kerr metric, because of the presence of the Carter constant, in Boyer-Lindquist coordinates the photon equations of motion are separable and of first order. Since the spacetime is stationary and axisymmetric, eventually it is only necessary to solve the motion in the $(r,\theta)$ plane, which (in the specific case of the Kerr metric in four dimensions) can be reduced to the problem of solving some elliptic integrals~\cite{cfm2}.

In extensions of the Kerr metric, the presence of the Carter constant still results in separable equations of motion, but one has now to solve at least hyper-elliptic integrals. In the case of a general stationary and axisymmetric metric without Carter constant, it is necessary to solve the second order geodesic equations. Usually it is computationally more convenient to start from the point of detection in the plane of the distant observer and trace backward in time the photon trajectory to find the emission point in the disk. More details can be found, for instance, in Ref.~\cite{c-cfm2}.

\begin{figure*}[t]
\begin{center}
\includegraphics[type=pdf,ext=.pdf,read=.pdf,width=8.5cm]{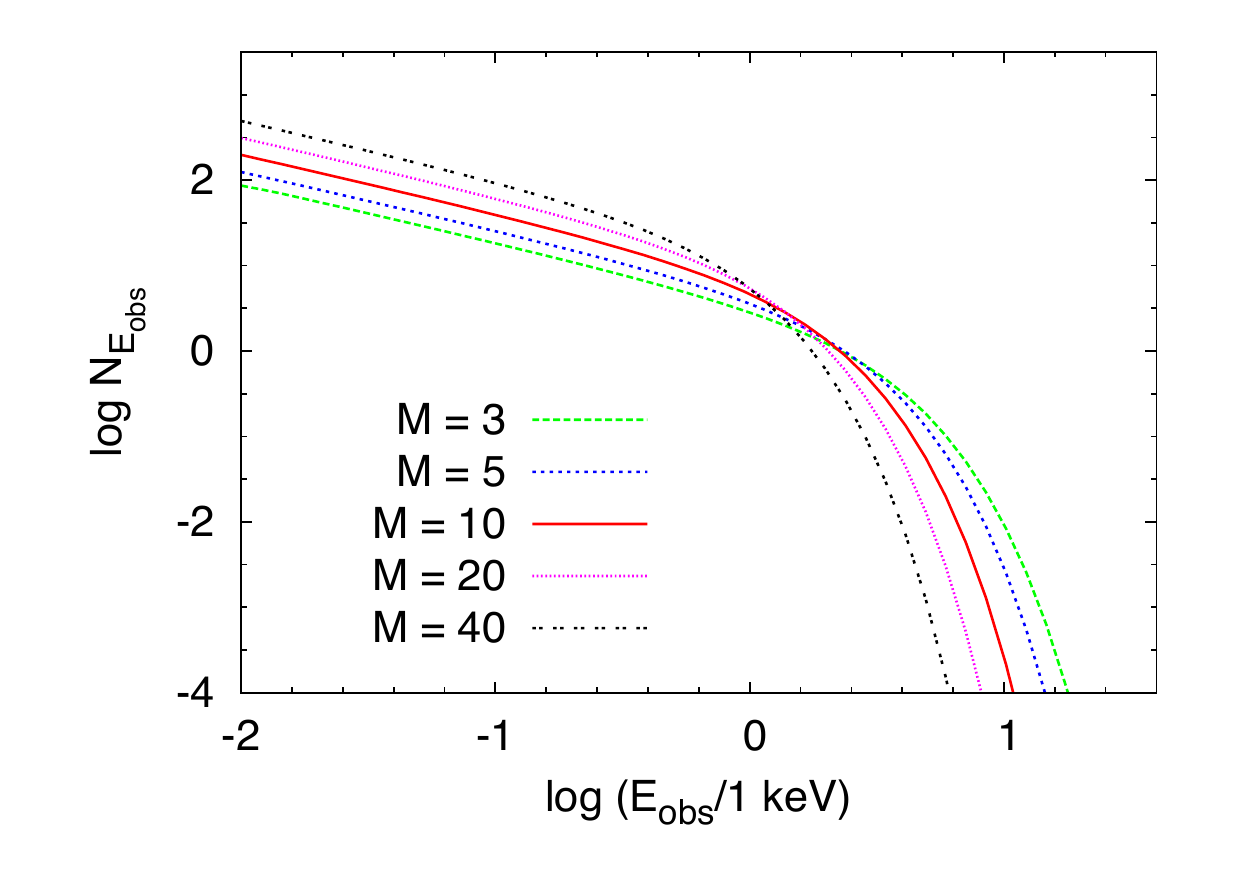}
\includegraphics[type=pdf,ext=.pdf,read=.pdf,width=8.5cm]{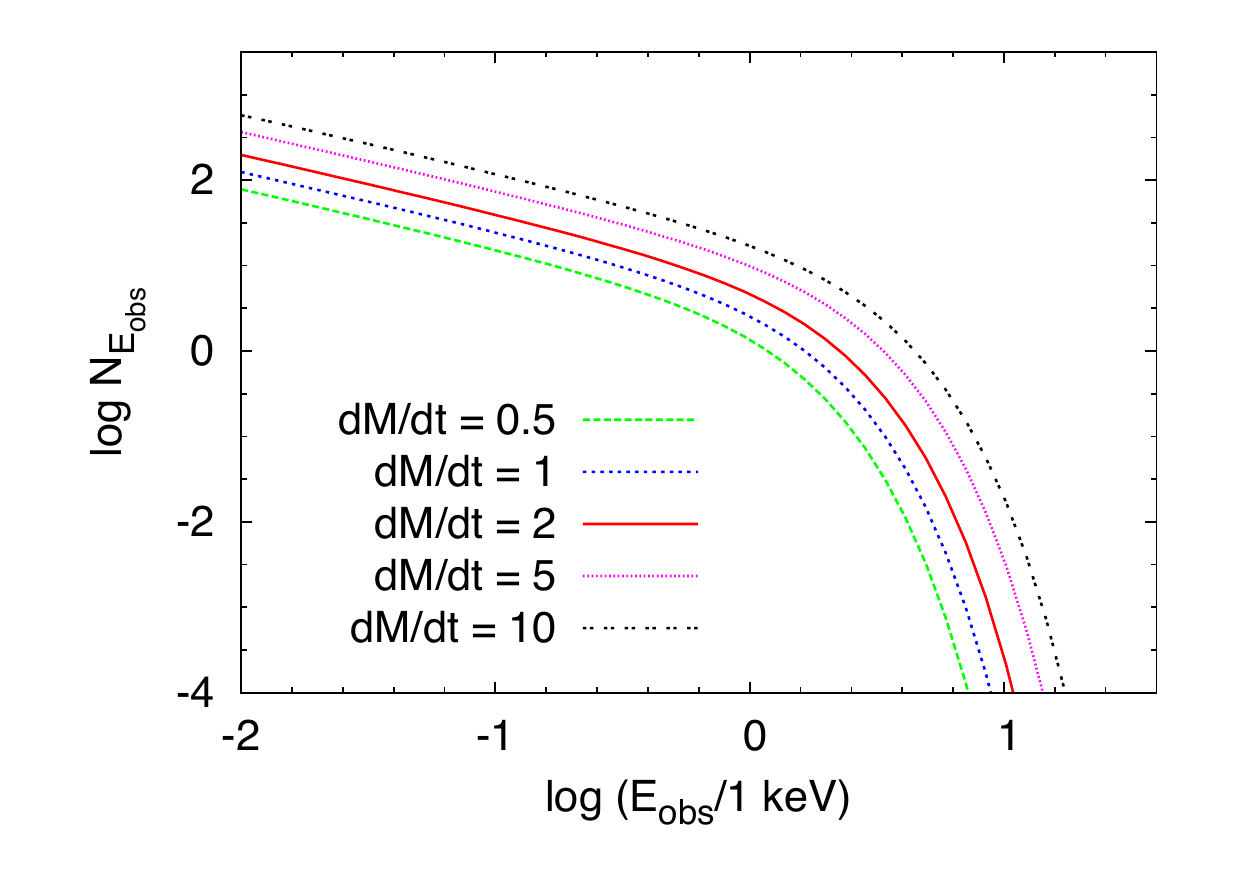} \\
\vspace{0.2cm}
\includegraphics[type=pdf,ext=.pdf,read=.pdf,width=8.5cm]{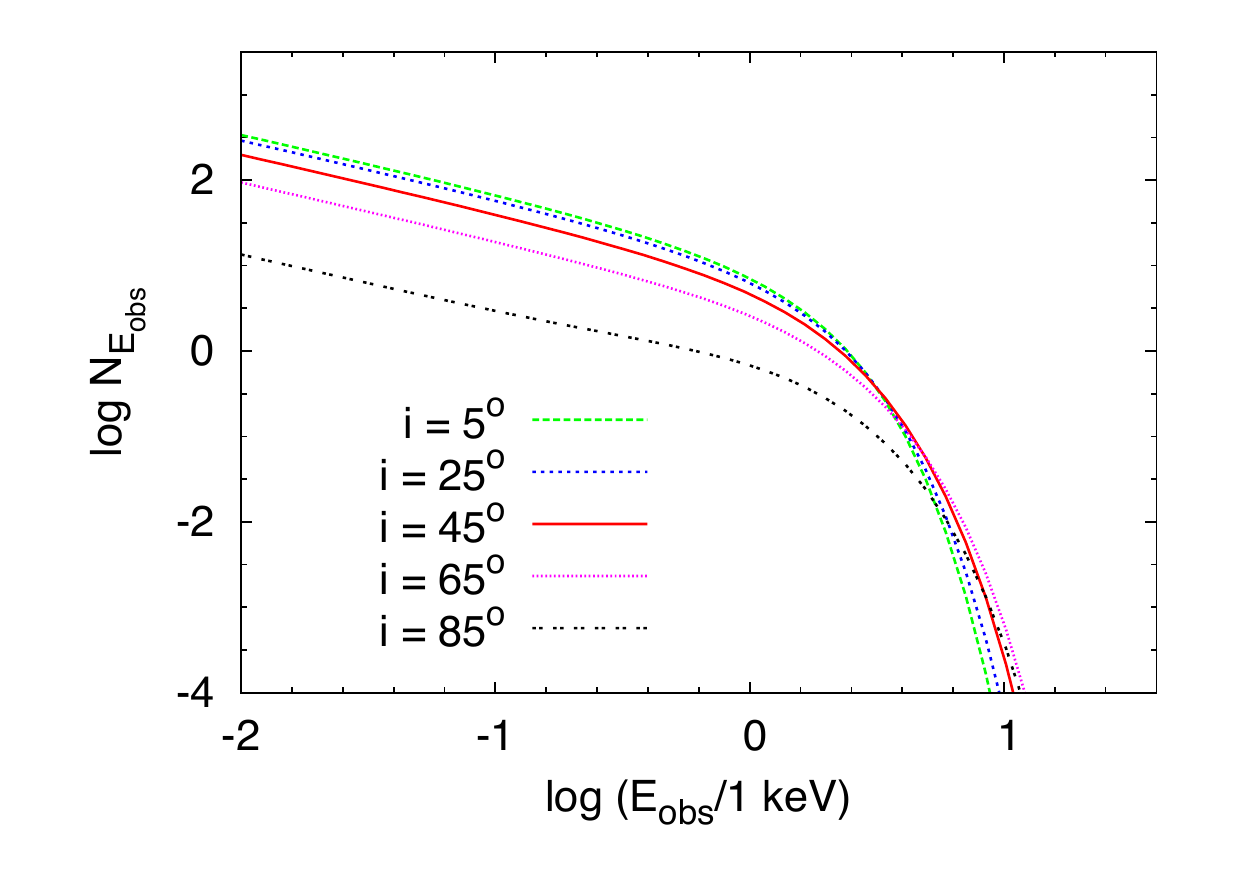}
\includegraphics[type=pdf,ext=.pdf,read=.pdf,width=8.5cm]{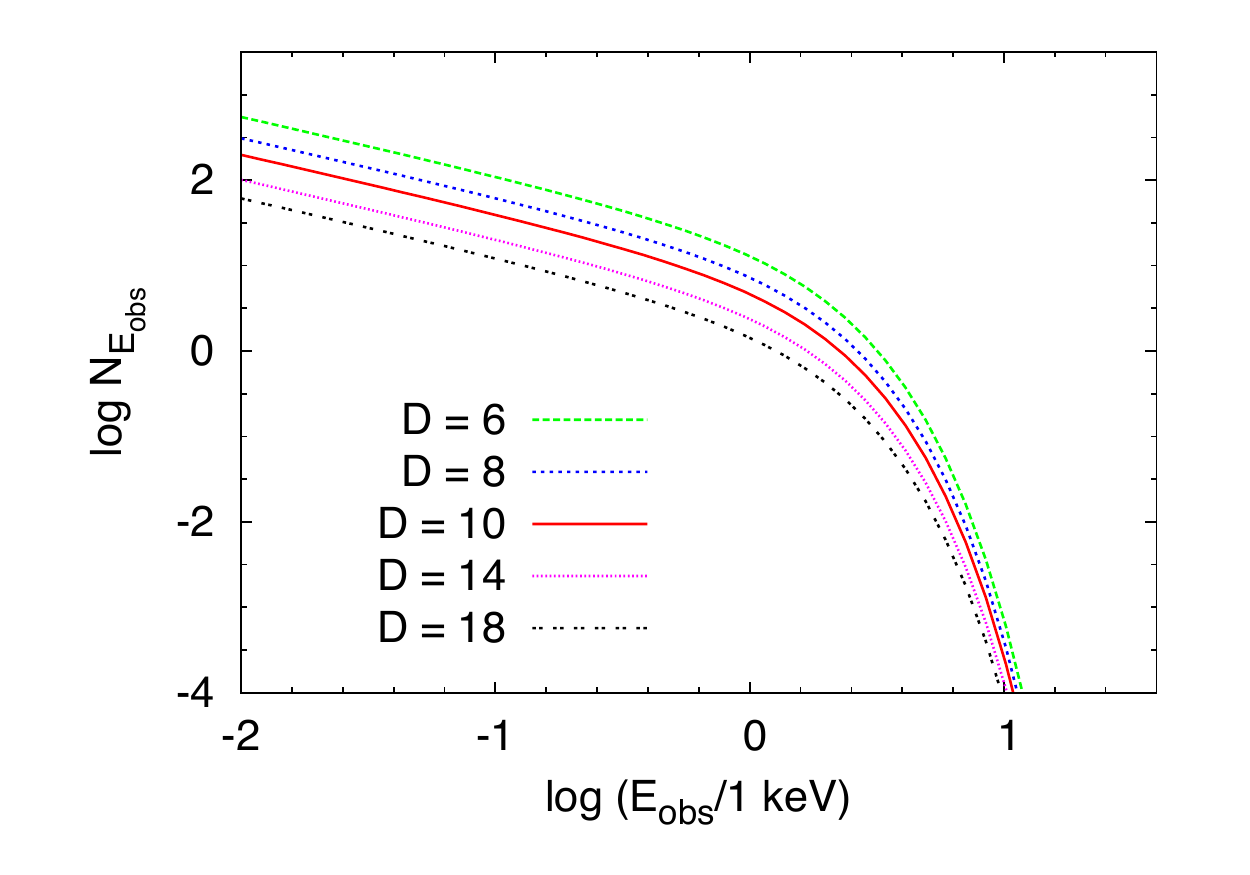} \\
\vspace{0.2cm}
\includegraphics[type=pdf,ext=.pdf,read=.pdf,width=8.5cm]{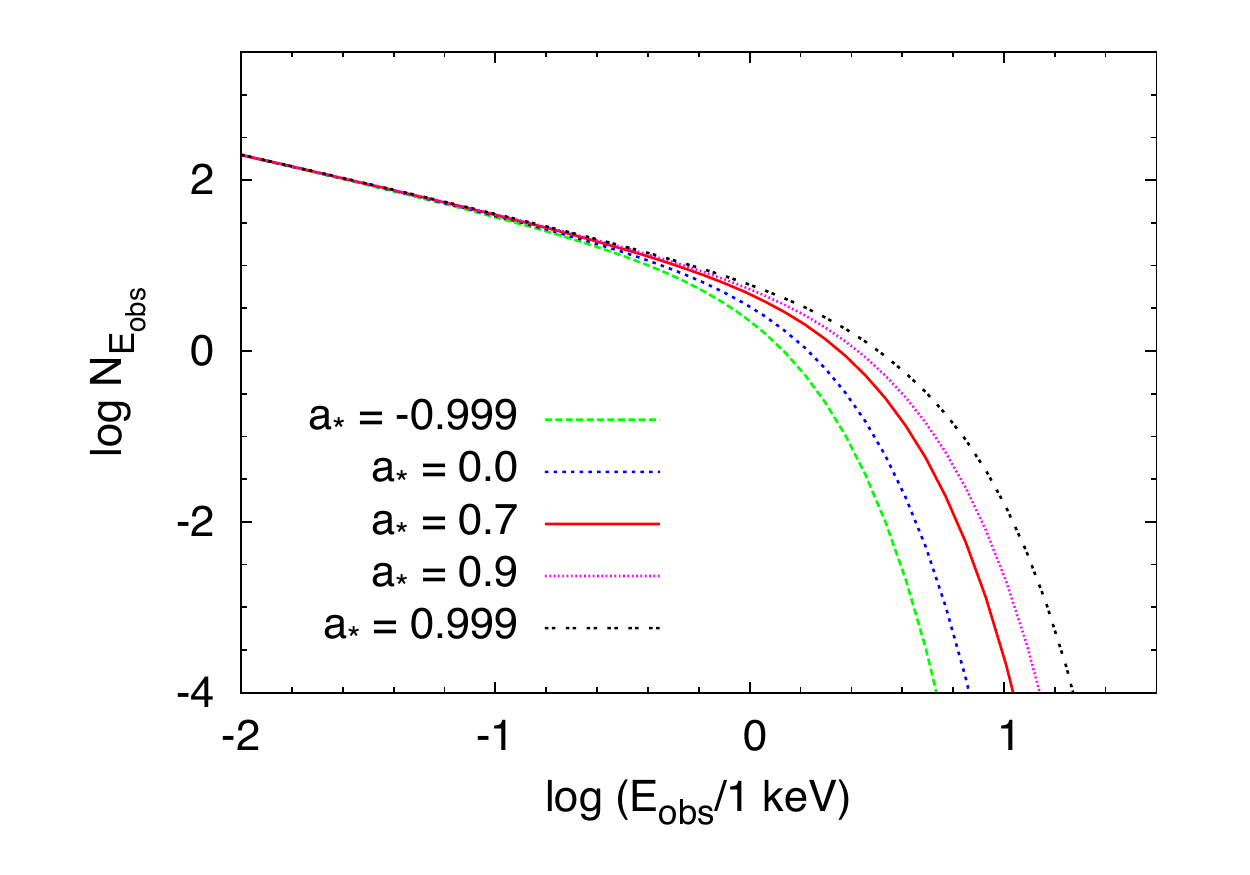}
\includegraphics[type=pdf,ext=.pdf,read=.pdf,width=8.5cm]{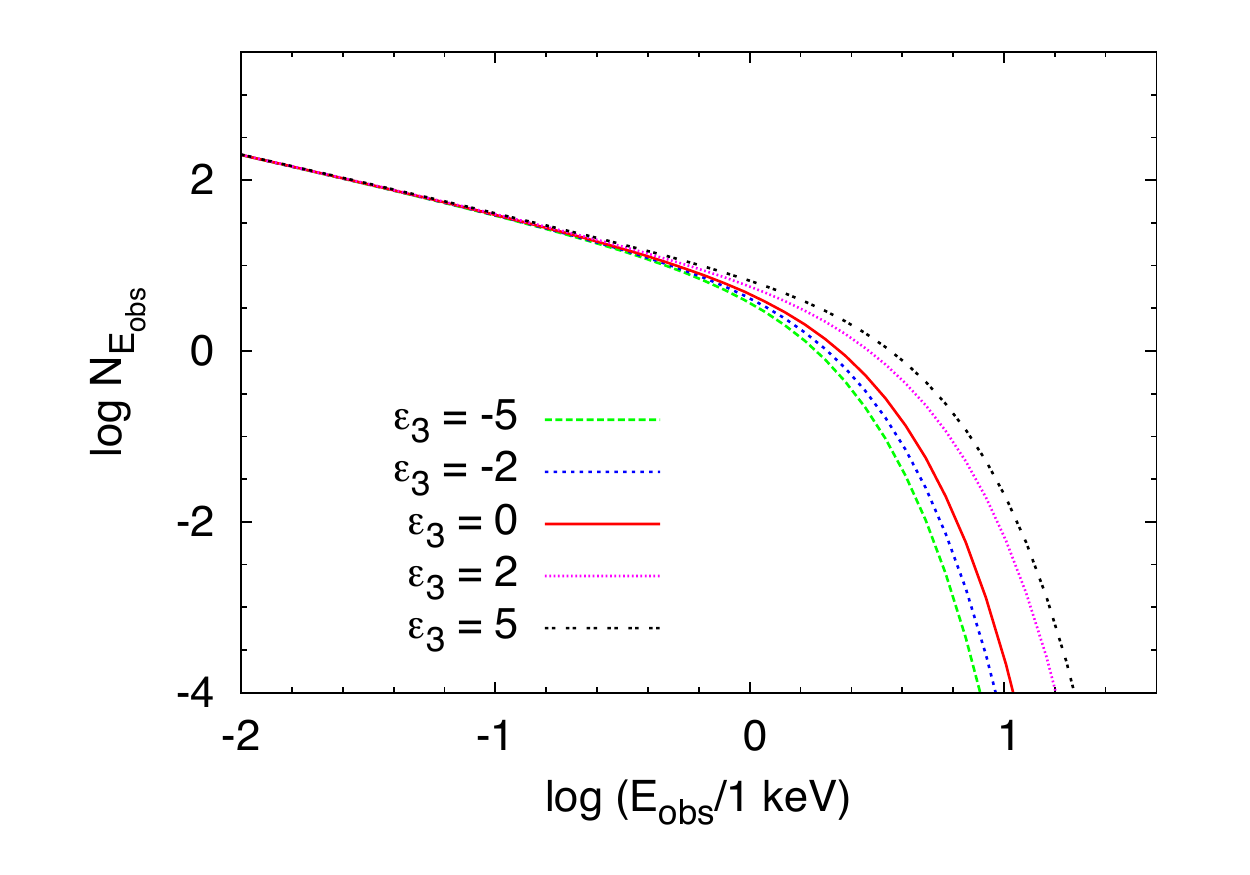} \\
\end{center}
\vspace{-0.2cm}
\caption{Impact of the model parameters on the thermal spectrum of a thin disk: mass $M$ (top left panel), mass accretion rate $\dot{M}$ (top right panel), viewing angle $i$ (central left panel), distance $D$ (central right panel), spin parameter $a_*$ (bottom left panel), and Johannsen-Psaltis deformation parameter $\epsilon_3$ (bottom right panel). When not shown, the values of the parameters are: $M = 10$~$M_\odot$, $\dot{M} = 2 \cdot 10^{18}$~g~s$^{-1}$, $D = 10$~kpc, $i = 45^\circ$, $a_*=0.7$, and $\epsilon_3 = 0$. $M$ in units of $M_\odot$, $\dot{M}$ in units of $10^{18}$~g~s$^{-1}$, $D$ in kpc, flux density $N_{E_{\rm obs}}$ in photons~keV$^{-1}$~cm$^{-2}$~s$^{-1}$, and photon energy $E_{\rm obs}$ in keV. }
\label{f-cfm1}
\end{figure*}

\begin{table*}[t]
\centering
\begin{tabular}{|ccccccc|}
\hline 
BH Binary & \hspace{0.1cm} & $a_*$ (CF) & \hspace{0.1cm} & $a_*$ (iron) & \hspace{0.1cm} & Principal References \\
\hline 
GRS~1915-105 && $> 0.98$ && $0.98 \pm 0.01$ && \cite{1915,1915b} \\
Cyg~X-1 && $> 0.98$ && $0.97^{+0.014}_{-0.02}$ && \cite{cyg1,cyg2,cyg3} \\
LMC~X-1 && $0.92 \pm 0.06$ && $0.97^{+0.02}_{-0.25}$ && \cite{lmcx1,lmcx1b} \\
GX~339-4 && $< 0.9$ && $0.95\pm0.03$ && \cite{gx339,gx339c} \\
MAXI~J1836-194 && --- && $0.88 \pm 0.03$ && \cite{maxi} \\
M33~X-7 && $0.84 \pm 0.05$ && --- && \cite{liu08} \\
4U~1543-47 && $0.80 \pm 0.10^\star$ && --- && \cite{sh06} \\
IC 10~X-1     &&  $\gtrsim0.7$  && --- && \cite{st16} \\
Swift~J1753.5 && --- && $0.76^{+0.11}_{-0.15}$ && \cite{swift} \\
XTE~J1650-500 && --- && $0.84 \sim 0.98$ && \cite{1650} \\
GRO~J1655-40 && $0.70 \pm 0.10^\star$ && $> 0.9$ && \cite{sh06,swift} \\
GS~1124-683 && $0.63^{+0.16}_{-0.19}$ && --- && \cite{gou_novamus} \\
XTE~J1752-223 && --- && $0.52 \pm 0.11$ && \cite{1752} \\
XTE~J1652-453 && --- && $< 0.5$ && \cite{1652} \\
XTE~J1550-564 && $0.34 \pm 0.28$ && $0.55^{+0.15}_{-0.22}$ && \cite{xte} \\
LMC~X-3 && $0.25 \pm 0.15$ && --- && \cite{lmcx3} \\
H1743-322 && $0.2 \pm 0.3$ && --- && \cite{h1743} \\
A0620-00 &&  $0.12 \pm 0.19$ && --- && \cite{62} \\
XMMU~J004243.6 && $< -0.2$ && --- && \cite{m31} \\
\hline 
\end{tabular}
\vspace{0.4cm}
\caption{Summary of the most recent continuum-fitting and iron line measurements of the spin parameter of stellar-mass black hole candidates under the assumption of the Kerr background. See the references in the last column for more details. $^\star$These sources were studied in an early work of the continuum-fitting method, within a more simple model, and therefore here the published 1-$\sigma$ error estimates are doubled. \label{tab1}}
\end{table*}

\begin{figure}[h]
\begin{center}
\includegraphics[type=pdf,ext=.pdf,read=.pdf,width=8.5cm]{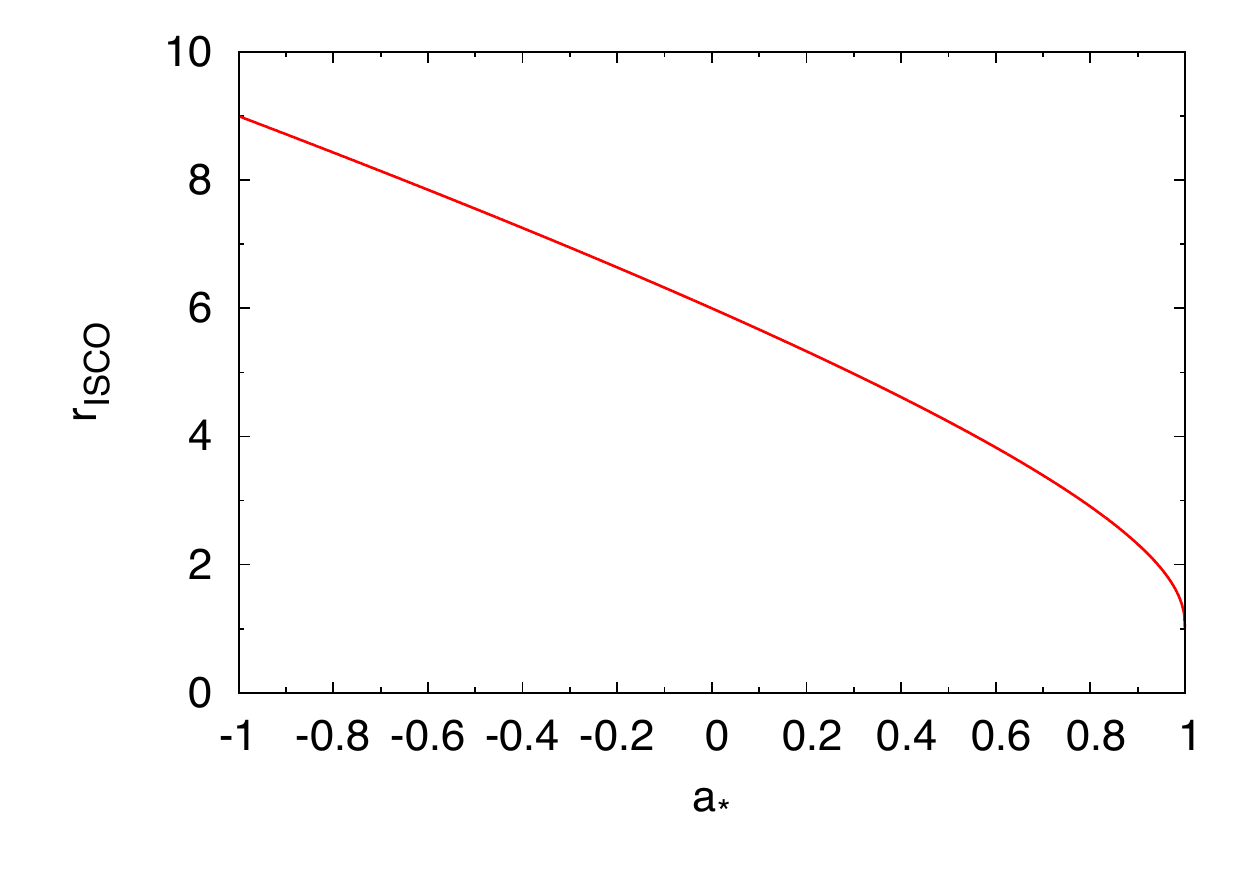}
\end{center}
\vspace{-0.2cm}
\caption{ISCO radius $r_{\rm ISCO}$ (in units $M=1$) versus the black hole spin parameter $a_*$ for the Kerr metric. }
\label{f-spin-isco}
\end{figure}

\subsection{Spin measurements with the continuum-fitting method \label{ss-cfm-s}}

The continuum-fitting method was originally proposed and then developed to measure the spin parameter under the assumption that astrophysical black hole candidates are the Kerr black holes of general relativity~\cite{cfm1,cfm3,cfm4}. Within the Novokov-Thorne model, the spectrum depends on five parameters: the black hole mass $M$, the black hole spin parameter $a_* = a/M$, the mass accretion rate $\dot{M}$, the inclination angle of the disk with respect to the line of sight of the distant observer $i$, and the distance of the source $D$. The impact of these parameters on the shape of the spectrum is shown in Fig.~\ref{f-cfm1}. If it is possible to get independent estimates of $M$, $i$, and $D$ (typically from optical observations), one can fit the soft X-ray component to infer $a_*$ and $\dot{M}$.

Current spin estimates via the continuum-fitting method of the spins of stellar-mass black hole candidates under the assumption of the Kerr background are reported in the second column in Tab.~\ref{tab1}. As we can see, some objects look like very fast-rotating black holes with $a_*$ close to 1, some objects have an intermediate value of the spin parameter, and other sources are consistent with non-rotating black holes. There is (controversial) evidence for at least one negative value of the spin parameter; that is, these objects would have a counterrotating disk.

The validity of the Novikov-Thorne model and therefore the reliability of the continuum-fitting method has been explored and confirmed in a number of studies. A requirement in the application of the method is that the source exhibits a strong contribution from the thermal disk emission.  Frequently, when employing models restricted to treating thin disks, one restrictively considers a luminosity range between a few percent and 30\% of the Eddington limit~\cite{1915}.

\begin{figure}
\begin{center}
\includegraphics[type=pdf,ext=.pdf,read=.pdf,width=8.0cm]{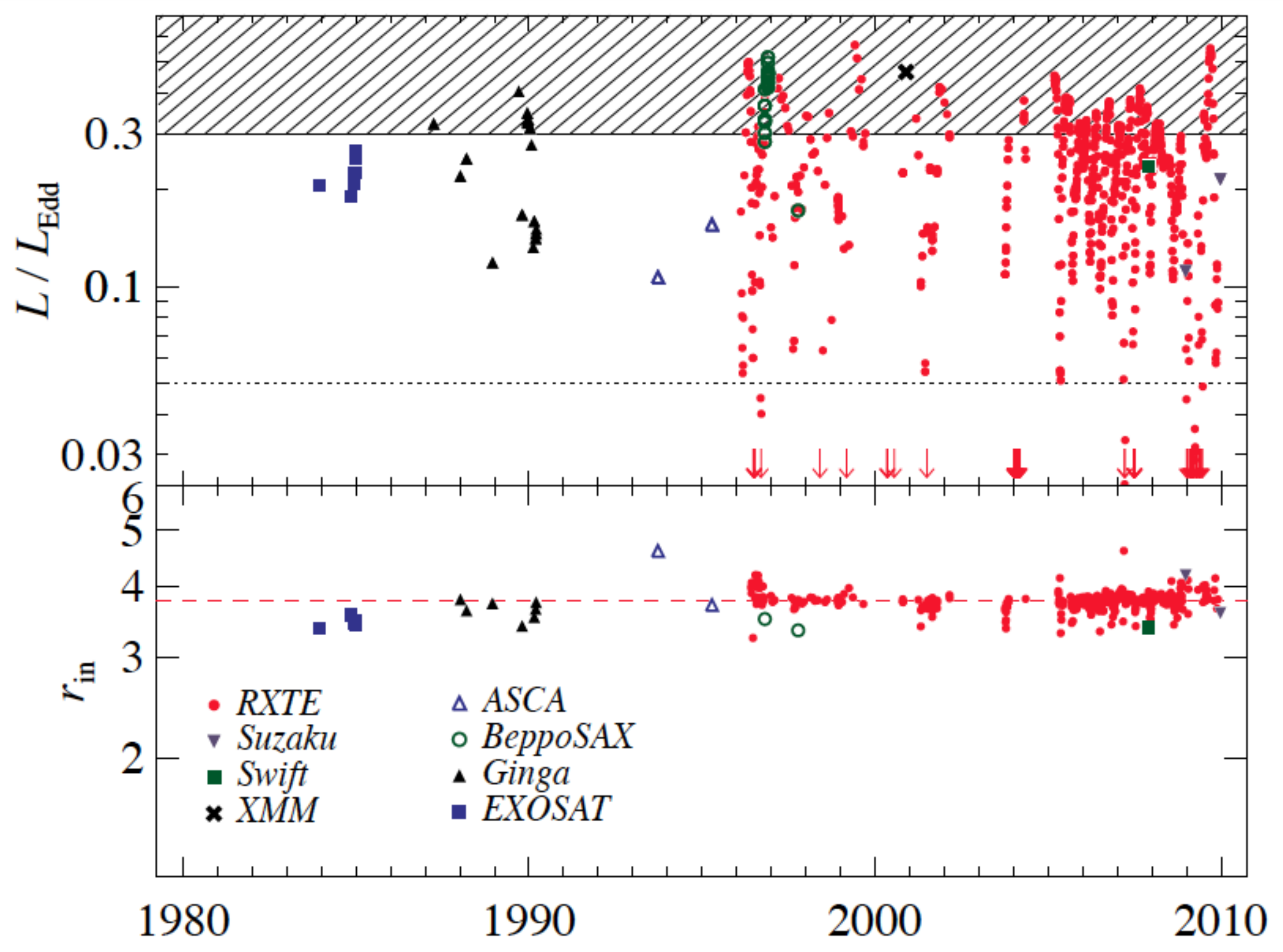}
\end{center}
\caption{Top panel: Accretion disk luminosity in Eddington units versus time for 766 spectra of LMC~X-3. The shaded region does not satisfy the thin disk selection criterion $L/L_{\rm Edd} < 0.3$, as well as the data below the dotted line, which marks $L/L_{\rm Edd} = 0.05$. Bottom panel: fitted value of the inner disk radius of the 411 spectra in the top panel can meet the thin disk selection criterion. From Ref.~\cite{constant}. See the text for more details. }
\label{f-steiner2010}
\vspace{0.8cm}
\begin{center}
\includegraphics[type=pdf,ext=.pdf,read=.pdf,width=8.0cm]{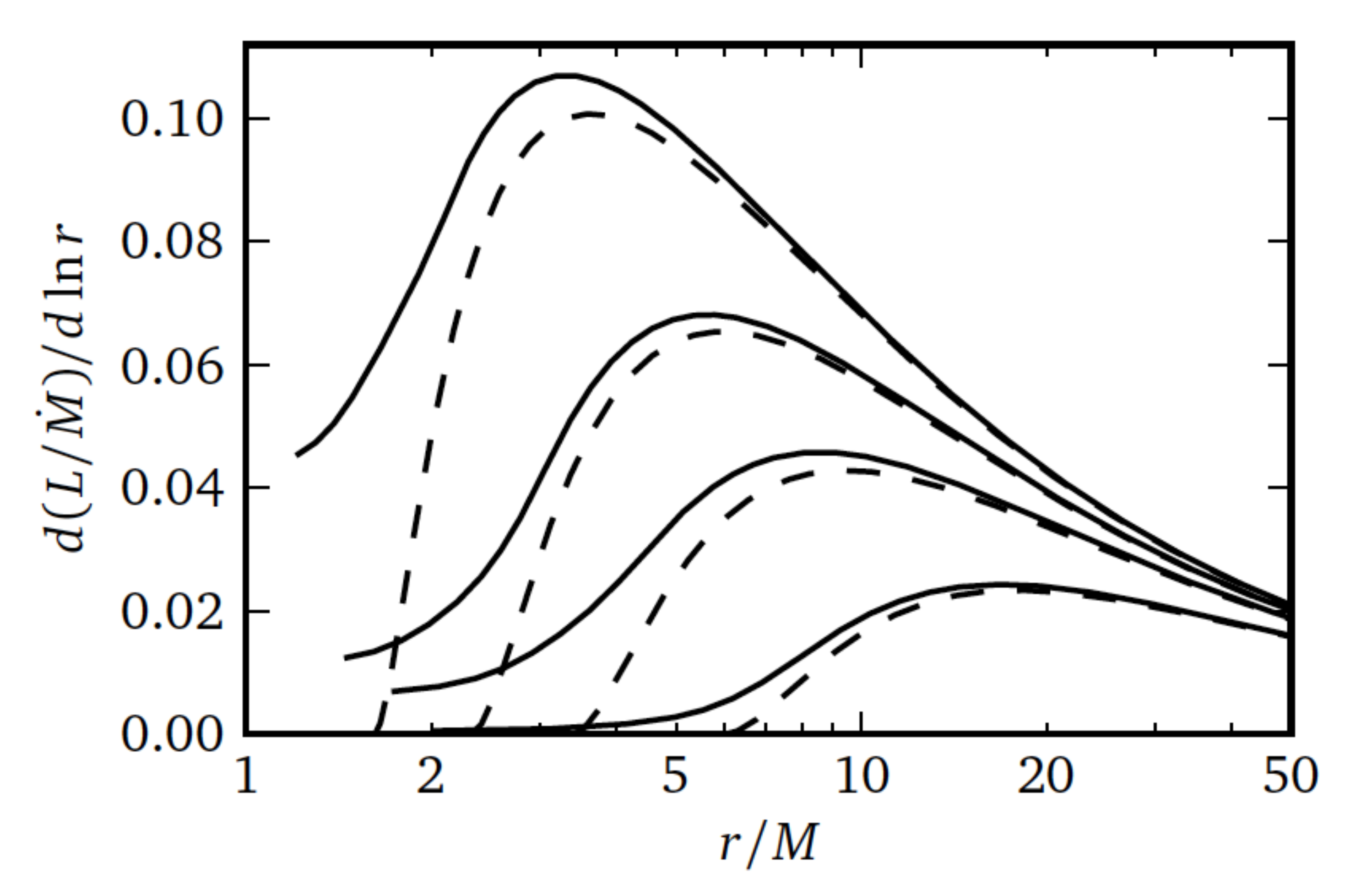}
\end{center}
\vspace{-0.2cm}
\caption{Luminosity profile from GRMHD simulations (solid lines) and from the Novikov-Thorne model (dashed lines) for $a_* = 0$, 0.7, 0.9, and 0.98 (from bottom to top). From Ref.~\cite{kulk10}. See the text for more details.}
\label{f-kulkarni2010}
\end{figure}

A crucial assumption in the model is that the inner edge of the disk is at the ISCO and that when the gas reaches the ISCO it quickly plunges onto the black hole without emitting other radiation. In the Kerr metric, there is a one-to-one correspondence between the spin parameter $a_*$ and the ISCO radius $r_{\rm ISCO}$ (see Fig.~\ref{f-spin-isco}) and this is the property which enables measurement of the spin. Observations show that the inner edge of the disk does not change appreciably over several years when the source is in the thermal state. The most compelling evidence comes from LMC~X-3. The analysis of many spectra collected during eight X-ray missions and spanning 26~years shows that the radius of the inner edge of the disk is quite constant~\cite{constant}, see Fig.~\ref{f-steiner2010}. The most natural interpretation is that the inner edge is associated to some intrinsic property of the geometry of the spacetime, namely the radius of the ISCO, and it is not affected by variable phenomena like the accretion process.

From the theoretical point of view, the fact that the inner edge of the disk is at the ISCO radius is related to the assumption of the model that the shear stress, driving the accretion at large radii, vanishes at the ISCO. Within a hydrodynamical description, the problem has been studied in Refs.~\cite{ntt1,ntt2,ntt3}, and the conclusion is that deviations from the Novikov-Thorne model decrease monotonically with the disk thickness $h/r$. So, thin disks with $h/r \ll 1$ should be well described by the Novikov-Thorne model. The case of magnetized accretion disks is more tricky, and studies with 3-dimensional general relativistic magnetohydrodynamics (GRMHD) simulations have been reported in Refs.~\cite{ntt4,ntt5,ntt6,ntt7,ntt8}. In particular, the impact of deviations from the Novikov-Thorne model in the spin measurements have been discussed in Ref.~\cite{kulk10}. Without invoking the details, it was found that there are indeed deviations from the theoretical model: some radiation is emitted inside the ISCO and the peak emission seems to be at smaller radii with respect to the Novikov-Thorne prediction. Both effects would lead to overestimate the value of the spin. Fig.~\ref{f-kulkarni2010} from Ref.~\cite{kulk10} compares the disk luminosity from GRMHD simulations (solid lines) with the Novikov-Thorne model (dashed lines) for $a_*=0$, 0.7, 0.9, 0.98 (from bottom to top). The error in the spin estimate due to these deviations from the Novikov-Thorne model is smaller for low viewing angles and high spins, and larger for high viewing angles and low spins. The deviations decrease as the disk thickness $h/r$ decreases. In the end, these effects do not seem to be important for current spin measurements, particularly  because spin errors are dominated by the uncertainties in the measurements of $M$, $D$, and $i$, but also because any additional radiation appears to manifest distinctly as the nonthermal component~\cite{kulk10, zhu12}.

Commonly for continuum fitting, one must make the assumption that the spin axis is aligned perpendicular to the binary orbital plane.  If the black hole candidate is the final product of the supernova explosion of a heavy star in a binary, its spin should be orthogonal to the orbital plane of the binary in the case of a symmetric explosion without strong shock and kick~\cite{fragos2010}. A misalignment may be introduced by a non-symmetric supernova explosion and/or shock and kick, as well as in those systems formed through multi-body interactions (binary capture or replacement), where the orientation of the spin of the black hole candidate and that of the orbital angular momentum of the binary are initially uncorrelated. However, the inner part of the disk -- which is the one important in the continuum-fitting measurement -- can be expected to be in the equatorial plane perpendicular to the spin of the black hole candidate as a result of the Bardeen-Petterson effect~\cite{bp,bp2}. This mechanism works for thin disks, because it requires $\alpha > h/r$, where $\alpha \sim 0.01 - 0.1$ is the viscosity parameter. The alignment timescale of thin disks should be $\sim 10^7$~years, and therefore the inner part of the disk should be in the equatorial plane if the black hole binary is not too young~\cite{6849} (but see~\cite{rm07m,rm08m} for more details). Future X-ray spectropolarimetric measurements of the thermal spectrum of the accretion disk will be able to check the validity of the assumption of the disk in the equatorial plane~\cite{lixin,schnittman,yifan}.

\begin{figure*}
\begin{center}
\includegraphics[type=pdf,ext=.pdf,read=.pdf,width=8.5cm]{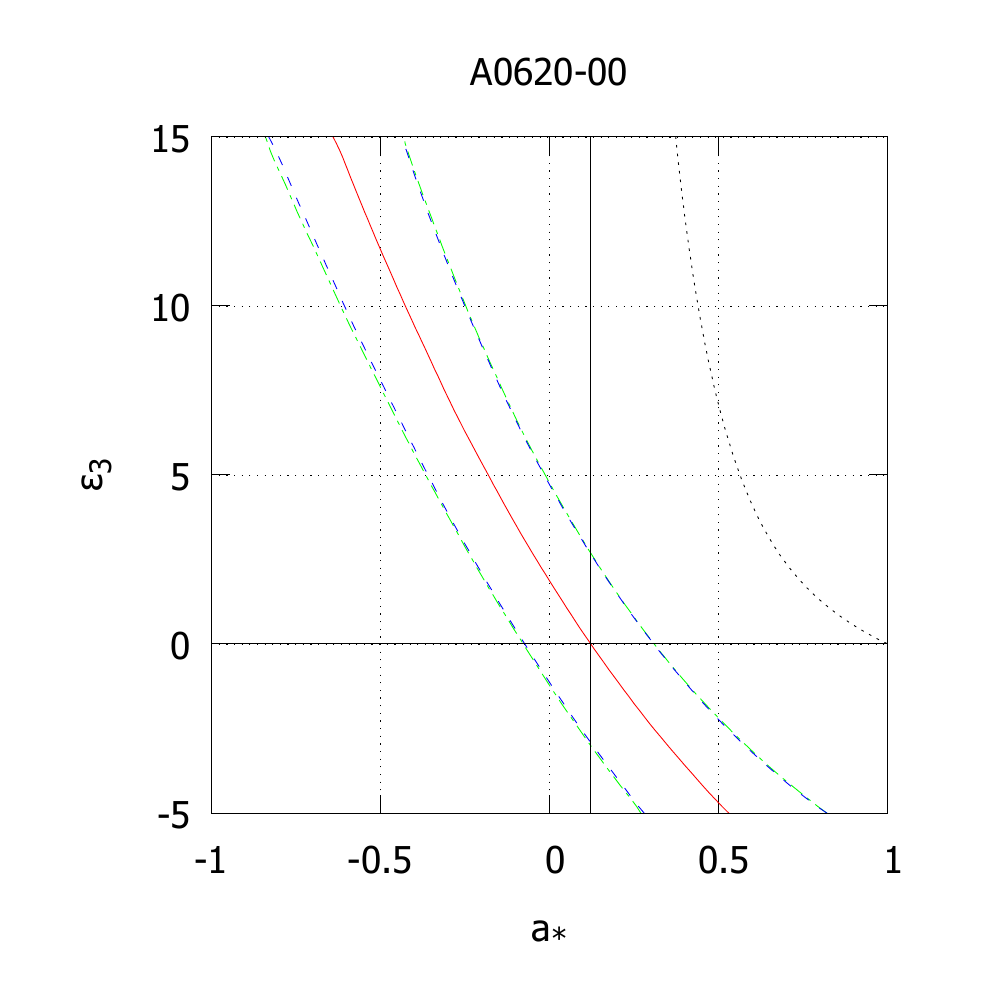}
\includegraphics[type=pdf,ext=.pdf,read=.pdf,width=8.5cm]{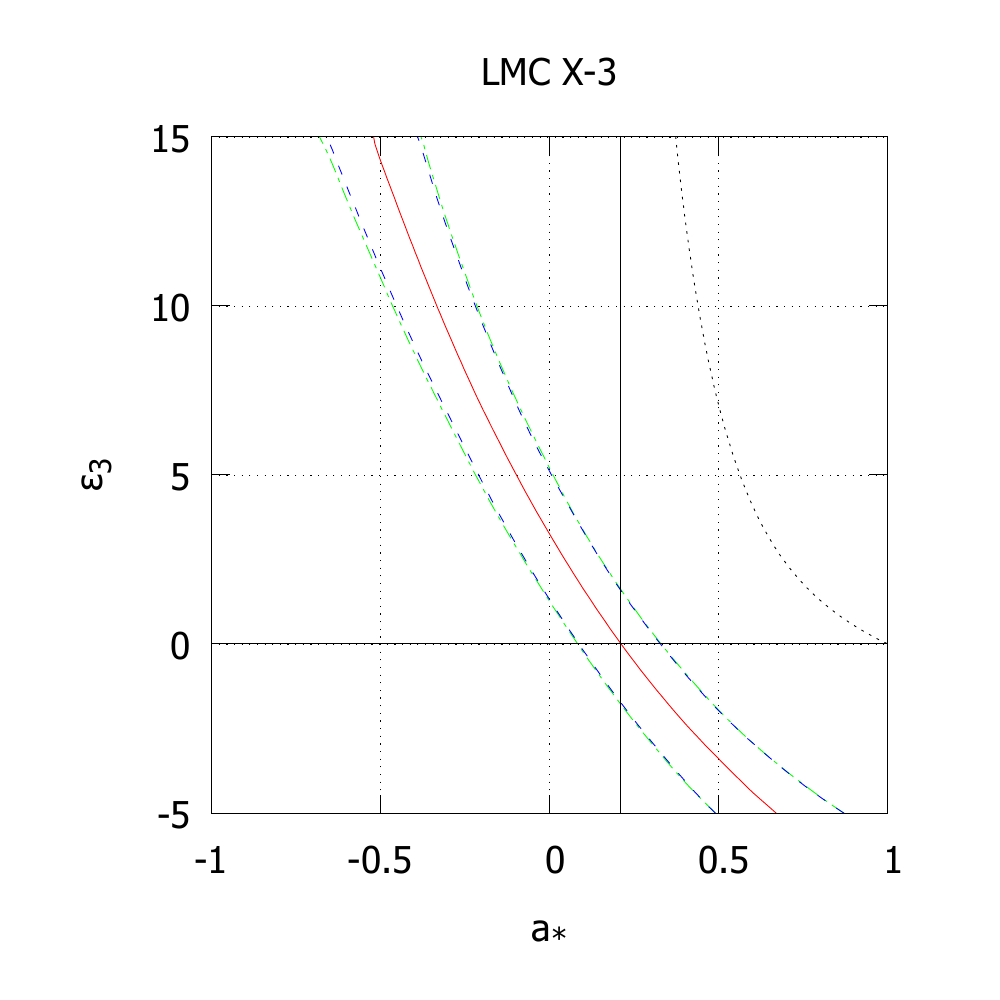} 
\end{center}
\vspace{-0.5cm}
\caption{Continuum-fitting constraints on the Johannsen-Psaltis parameter $\epsilon_3$ from A0620-00 (left panel) and LMC~X-3 (right panel). The black dotted curve starting from the point $a_*=1$ and $\epsilon_3 = 0$ and moving to lower values of the spin parameter at larger $\epsilon_3$ separates the region in which the horizon completely covers the black hole singularity (on the left of the dotted curve) from the region in which the black hole singularity is naked (on the right of the dotted curve). From Ref.~\cite{lingyao}. See the text for more details. }
\label{f-cfm2}
\vspace{0.5cm}
\begin{center}
\includegraphics[type=pdf,ext=.pdf,read=.pdf,width=8.5cm]{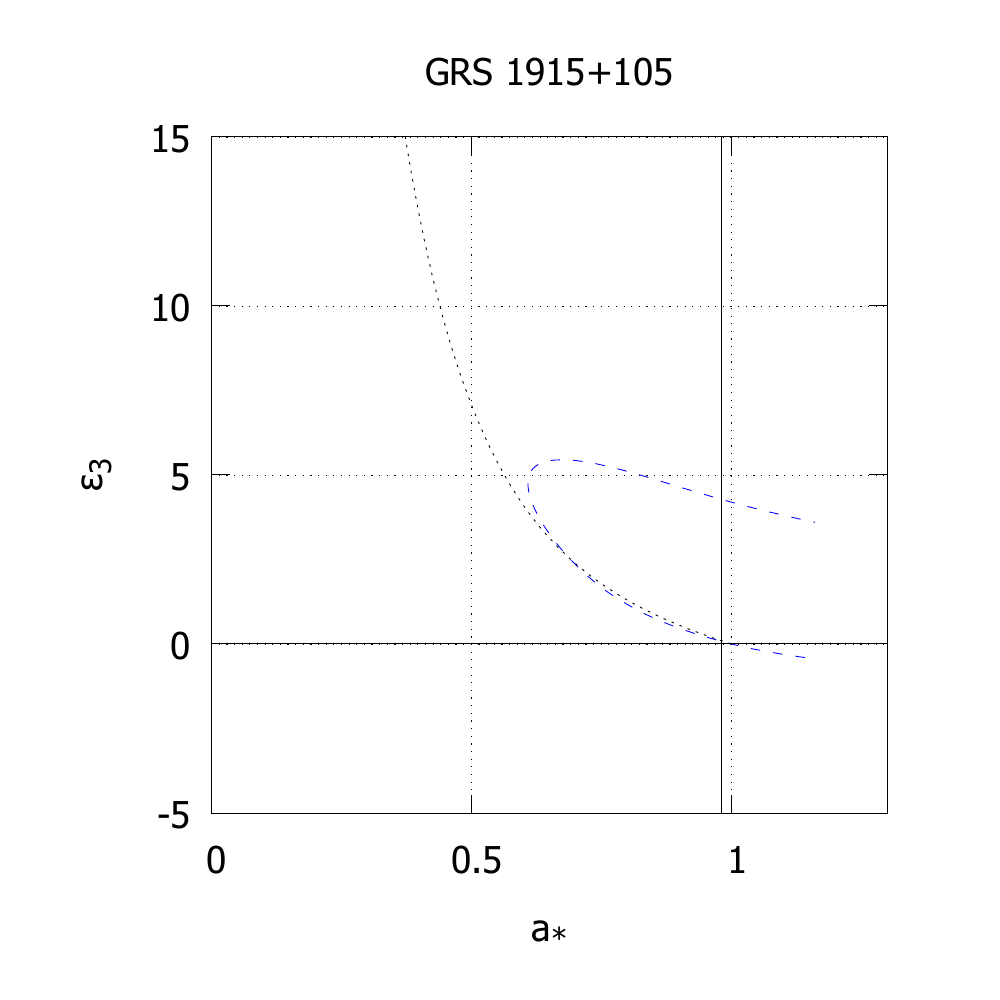}
\put(-93,102){Allowed Region}
\put(-160,145){Excluded Region}
\includegraphics[type=pdf,ext=.pdf,read=.pdf,width=8.5cm]{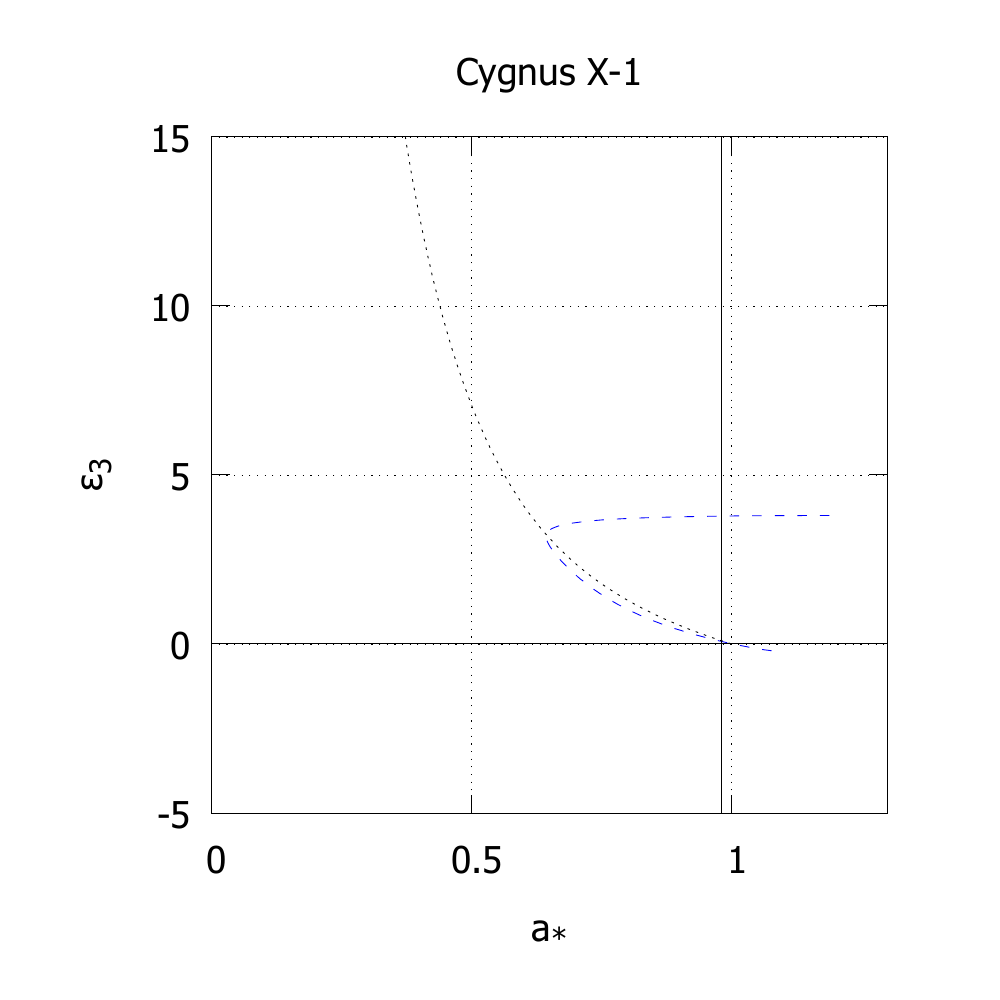}
\put(-93,102){Allowed Region}
\put(-160,145){Excluded Region}
\end{center}
\vspace{-0.5cm}
\caption{Continuum-fitting constraints on the Johannsen-Psaltis parameter $\epsilon_3$ from GRS~1915+105 (left panel) and Cygnus~X-1 (right panel). The black dotted curve starting from the point $a_*=1$ and $\epsilon_3 = 0$ and moving to lower values of the spin parameter at larger $\epsilon_3$ separates the region in which the horizon completely covers the black hole singularity (on the left of the dotted curve) from the region in which the black hole singularity is naked (on the right of the dotted curve). From Ref.~\cite{lingyao}. See the text for more details.}
\label{f-cfm3}
\end{figure*}

\subsection{Testing black hole candidates with the continuum-fitting method \label{ss-cfm}}

The Novikov-Thorne model is formulated in a generic stationary, axisymmetric, and asymptotically flat spacetime. The generalization of the continuum-fitting technique to non-Kerr background is thus quite straightforward. The calculations of the spectrum are done in a different way, because it is not possible to exploit the nice properties of the Kerr metric, in which the resolution of the equations of motion can be reduced to a calculation of elliptic integrals. The details of the calculations for non-Kerr metrics can be found in~\cite{c-cfm2}. It is also worth noting that in the Kerr metric the orbits in the equatorial plane are always vertically stable and the ISCO is determined by the stability along the radial direction. In non-Kerr backgrounds, this may not be true and the ISCO radius may be marginally vertically stable.  Additionally, whereas crossing the ISCO radius is sufficient for freefall in Kerr, more generally, there may be spacetimes for which the gas cannot plunge from the ISCO to the black hole, but it must lose additional energy and angular momentum~\cite{isco}.

The thermal spectrum of thin accretion disks in non-Kerr spacetimes has been studied in~\cite{t-cfm,h-cfm,h-cfm1,h-cfm2,h-cfm3,c-cfm1,c-cfm2b,lingyao,c-cfm3}. Ref.~\cite{lingyao} revises the current spin measurements obtained with the assumption of the Kerr background in the case of the Johannsen-Psaltis metric with non-vanishing deformation parameter $\epsilon_3$. Figs.~\ref{f-cfm2} and~\ref{f-cfm3} show some representative constraints from the continuum-fitting method on the parameter $\epsilon_3$\footnote{We note that, while in the Kerr metric it is possible to restrict the study to $|a_*| \le 1$ (see the discussion in Ref.~\cite{review-l} for the details), in non-Kerr backgrounds one should consider also objects with $|a_*| > 1$, which may still be black holes and they may be created by physically reasonable mechanisms. The spin-up of Manko-Novikov compact objects (which are not black holes because they may have singularities and regions with closed time-like curves outside the event horizon) is discussed in Refs.~\cite{spin1,spin2}. The same mechanism should work for some non-Kerr black holes.}.

In the case of objects that look like slow-rotating Kerr black holes, i.e., sources with low Novikov-Thorne radiative efficiency $\eta_{\rm NT} = 1 - E_{\rm ISCO}$, where $E_{\rm ISCO}$ is the specific energy of a test-particle at the ISCO, there is a measurement degeneracy between the spin parameter $a_*$ and the deformation parameter $\epsilon_3$. The Novikov-Thorne radiative efficiency corresponds to the constant relating the accretion luminosity $L_{\rm acc}$ and the mass accretion rate $\dot{M}$ through the disk, namely $L_{\rm acc} = \eta_{\rm NT} \dot{M}$. A low $\eta_{\rm NT}$ is the case, for instance, for the black hole candidates in A0620-00 and LMC~X-3 shown in Fig.~\ref{f-cfm2}. The parameter settings along the red line are indistinguishable and the spin estimate within the Kerr hypothesis becomes an allowed region in the spin parameter-deformation parameter plane (the regions between the two blue dashed lines in Figs.~\ref{f-cfm2}).

If a source looks like a fast-rotating Kerr black hole, namely it has a high Novikov-Thorne radiative efficiency $\eta_{\rm NT}$, then it is possible to constrain $\epsilon_3$. This is the case for GRS1915+105 and Cygnus~X-1 shown in Fig.~\ref{f-cfm3}. The same can be done in the case of other deformation parameters, see e.g. Ref.~\cite{c-cfm2b}, but not with all~\cite{c-cfm3}. The point is that very deformed objects may not be able to have a very high radiative efficiency. In other words, it is not always easy to mimic very fast-rotating Kerr black holes.

\begin{figure*}
\begin{center}
\includegraphics[type=pdf,ext=.pdf,read=.pdf,width=8.5cm]{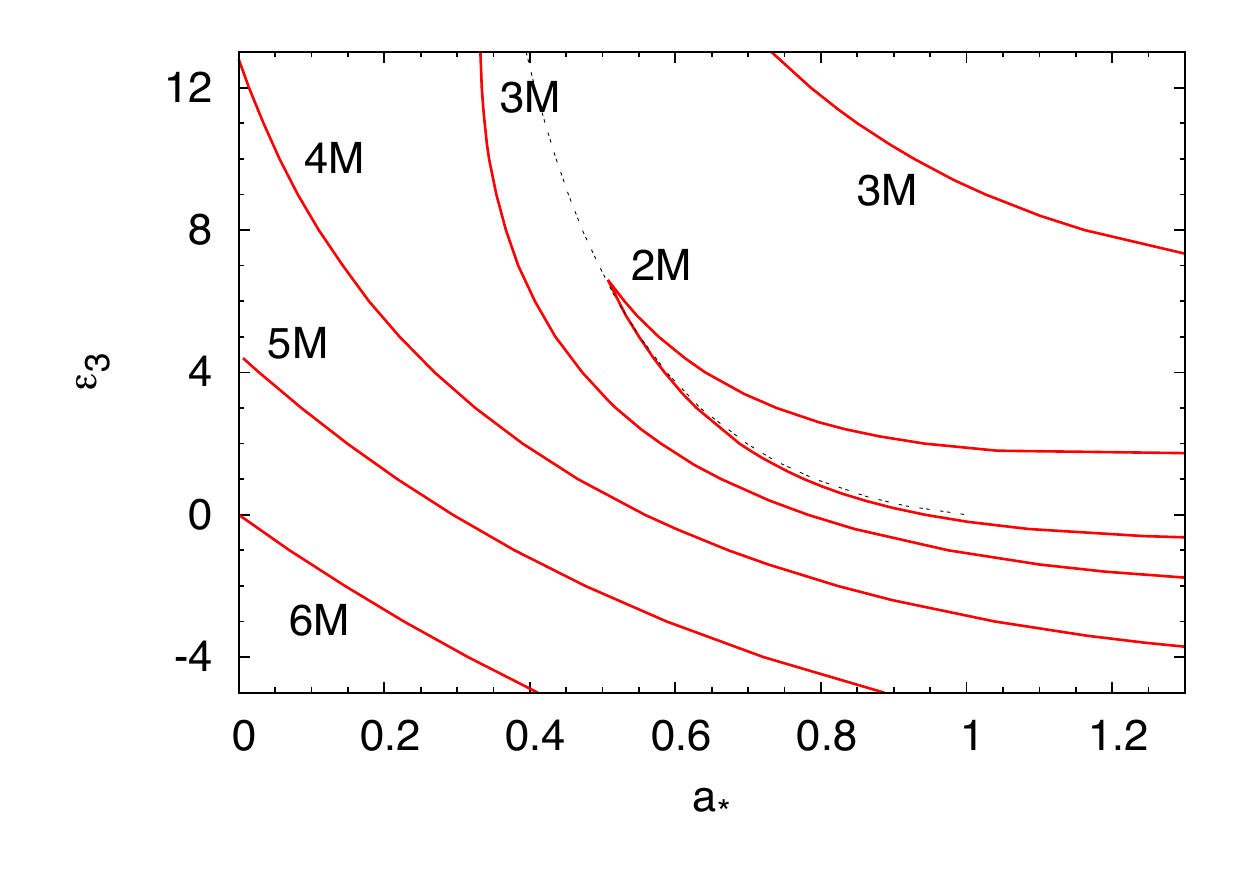}
\includegraphics[type=pdf,ext=.pdf,read=.pdf,width=8.5cm]{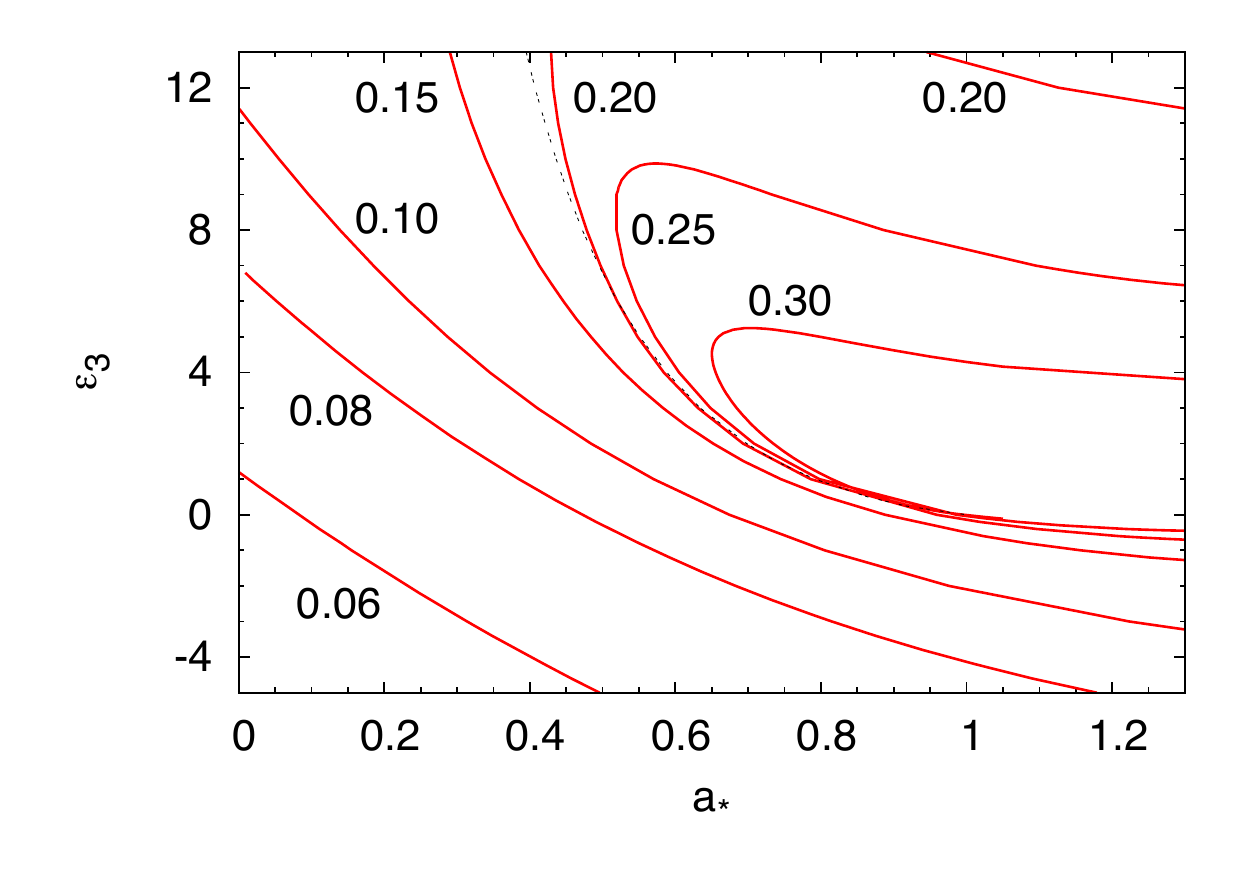}
\end{center}
\vspace{-0.5cm}
\caption{Contour levels of the ISCO radius $r_{\rm ISCO}$ (left panel) and of the Novikov-Thorne radiative efficiency $\eta_{\rm NT} = 1 - E_{\rm ISCO}$ (right panel) in the Johannsen-Psaltis metric with non-vanishing deformation parameter $\epsilon_3$. The black dotted line separates spacetime with a regular event horizon (on the left of the line) from those with topologically non-trivial event horizons and naked singularities (on the right). See the text for more details. }
\label{f-cfm4}
\end{figure*}

In the Kerr metric, there is a one-to-one correspondence between the spin parameter $a_*$ and the ISCO radius $r_{\rm ISCO}$, or the Novikov-Thorne radiative efficiency $\eta_{\rm NT}$. This is the key observable that permits the estimate of the spin.  $r_{\rm ISCO}$, or $\eta_{\rm NT}$ become degenerate in the presence of one more parameter of the spacetime geometry which perturbs the Kerr solution. Actually, in a non-Kerr metric it is possible to see that one is most sensitive to the radiative efficiency. Fig.~\ref{f-cfm4} shows the contour levels of $r_{\rm ISCO}$ (left panel) and $\eta_{\rm NT}$ (right panel) in the plane spin parameter $a_*$ versus Johannsen-Psaltis deformation parameter $\epsilon_3$. The shape of the contour lines of $\eta_{\rm NT}$ is very similar to the shape of the constraints in Fig.~\ref{f-cfm2} and Fig.~\ref{f-cfm3} (see Ref.~\cite{lingyao} for more examples).

Lastly, we note that, in principle, an observation may identify a black hole candidate with a Novikov-Thorne radiative efficiency too high to be explainable as a Kerr black hole. For a Kerr black hole, the maximum value of $\eta_{\rm NT}$ is about 0.42 and is reached for $a_* = 1$. Such an observation would be an indication of new physics. Notably, all objects studied have been within the Kerr bounds; all black hole candidates studied with the continuum-fitting method are consistent with the Kerr black hole hypothesis.

\section{Iron line spectroscopy}

The X-ray spectrum of both stellar-mass and supermassive black hole candidates presents emission lines that are interpreted as the result of the illumination of a cold accretion disk by a hot corona. One of the most prominent emission lines is  iron K$\alpha$  at 6-4-7.0~keV, depending on ionization state: this line is intrinsically narrow in frequency, while the one in the spectrum of black hole candidates becomes broad and skewed due to relativistic effects. Other explanations have been proposed, but reverberation lag measurements provide strong support of the relativistic  interpretation and disfavor or rule out other explanations~\cite{rev-rev}.

While the continuum-fitting method requires independent measurements of the mass, the distance, and the viewing angle of the source, the iron line analysis does not need them: mass and distance have no impact on the line profile, as the physics is essentially independent of the size of the system, while the viewing angle can be inferred during the fitting procedure from the effect of the Doppler blueshift.

\subsection{Disk-corona model}

\begin{figure}[b]
\begin{center}
\includegraphics[type=pdf,ext=.pdf,read=.pdf,width=8.5cm]{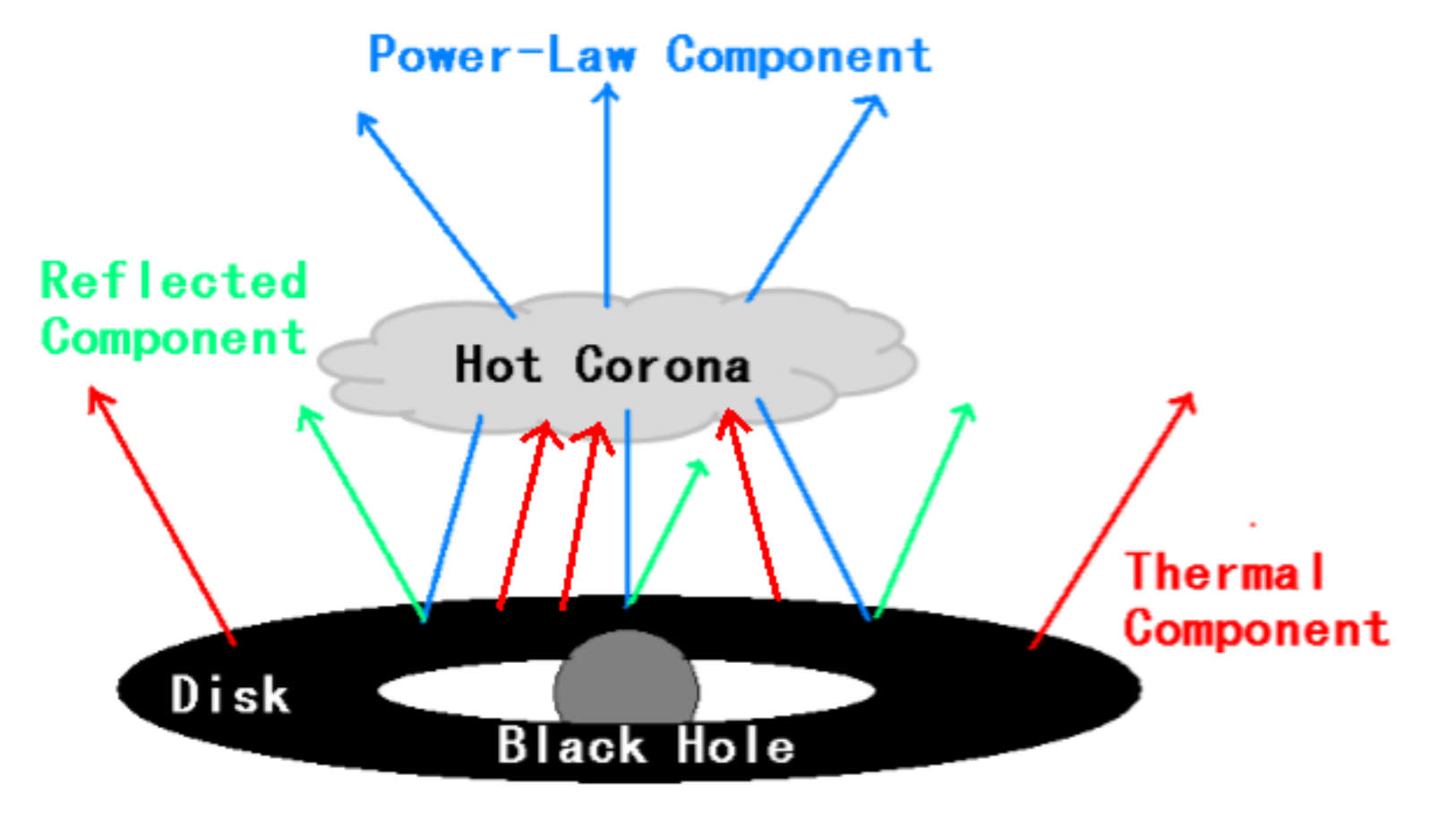}
\end{center}
\vspace{-0.5cm}
\caption{Corona-disk model with a schematic configuration similar to the widely-used lamppost geometry in the case of vanishing coronal size. The black hole has a geometrically thin and optically thick accretion disk. The corona is a hot electron cloud acting as an X-ray source and located just above the black hole. The corona illuminates  the disk, producing a reflection component, the most prominent feature of which is usually the iron K$\alpha$ line. }
\label{f-diskcorona}
\end{figure}

The actual geometry of the corona is not known and several scenarios may be possible. A configuration similar to the so-called lamppost geometry is sketched in Fig.~\ref{f-diskcorona}. Here we have a black hole surrounded by a cold thin accretion disk and there is a ``hot corona'' just above the black hole. Within this set-up, the corona may be the base of a jet. In the corona, there are hot electrons cooled by inverse Compton scattering from the thermal spectrum of the disk. The hot corona's scattered photons act as an X-ray source, which is seen as a power-law component in the X-ray spectrum of the object. Some radiation illuminates  the disk, producing a reflected component which includes emission lines, the most prominent of which is the iron K$\alpha$ line.

As in the continuum-fitting method, the iron line measurement assumes that the accretion disk can be described by the Novikov-Thorne model~\cite{ntm,ntm2}, and  crucially depends on the inner edge of the disk being located at the ISCO radius. The latter point can be tricky for the iron line analysis, because there is controversy over this assumption in hard states which are those most frequently employed in reflection modeling.  Specifically, it is argued whether the disk is in fact truncated at some radius larger than the ISCO~\cite{v2-isco1,v2-isco2,v2-isco3}. At present, evidence indicates that if truncation is present, it is likely to be mild (a factor of a few).  If one were to fit data for a system in which the disk were truncated, and make the usual assumption that the inner-edge was at the ISCO, then the fit would incorrectly underestimate the value of spin (relative to the resulting Kerr prediction if it were truncated at the ISCO radius).  The same question is at play for using reflection to test the Kerr metric with data from faint hard-states.

\begin{figure*}
\begin{center}
\includegraphics[type=pdf,ext=.pdf,read=.pdf,width=8.5cm]{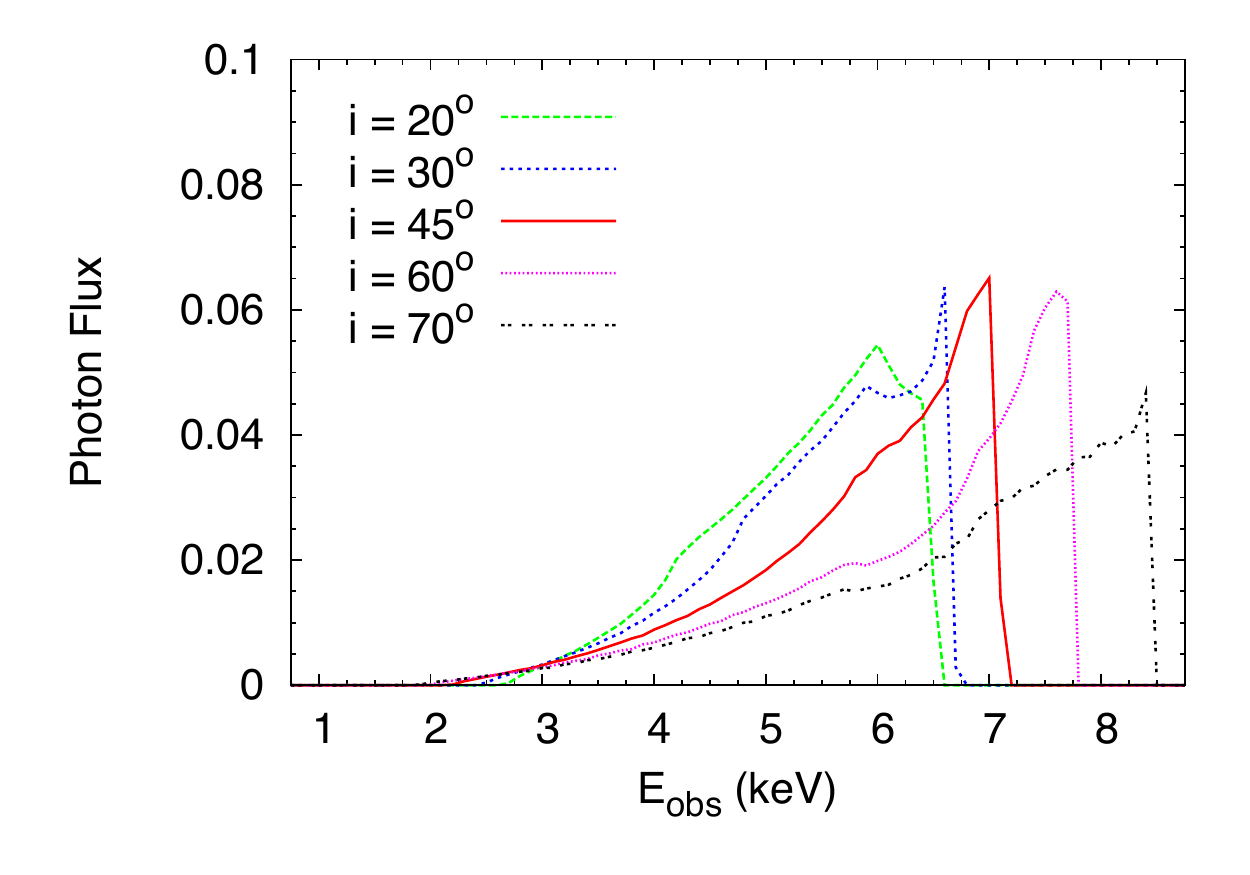}
\includegraphics[type=pdf,ext=.pdf,read=.pdf,width=8.5cm]{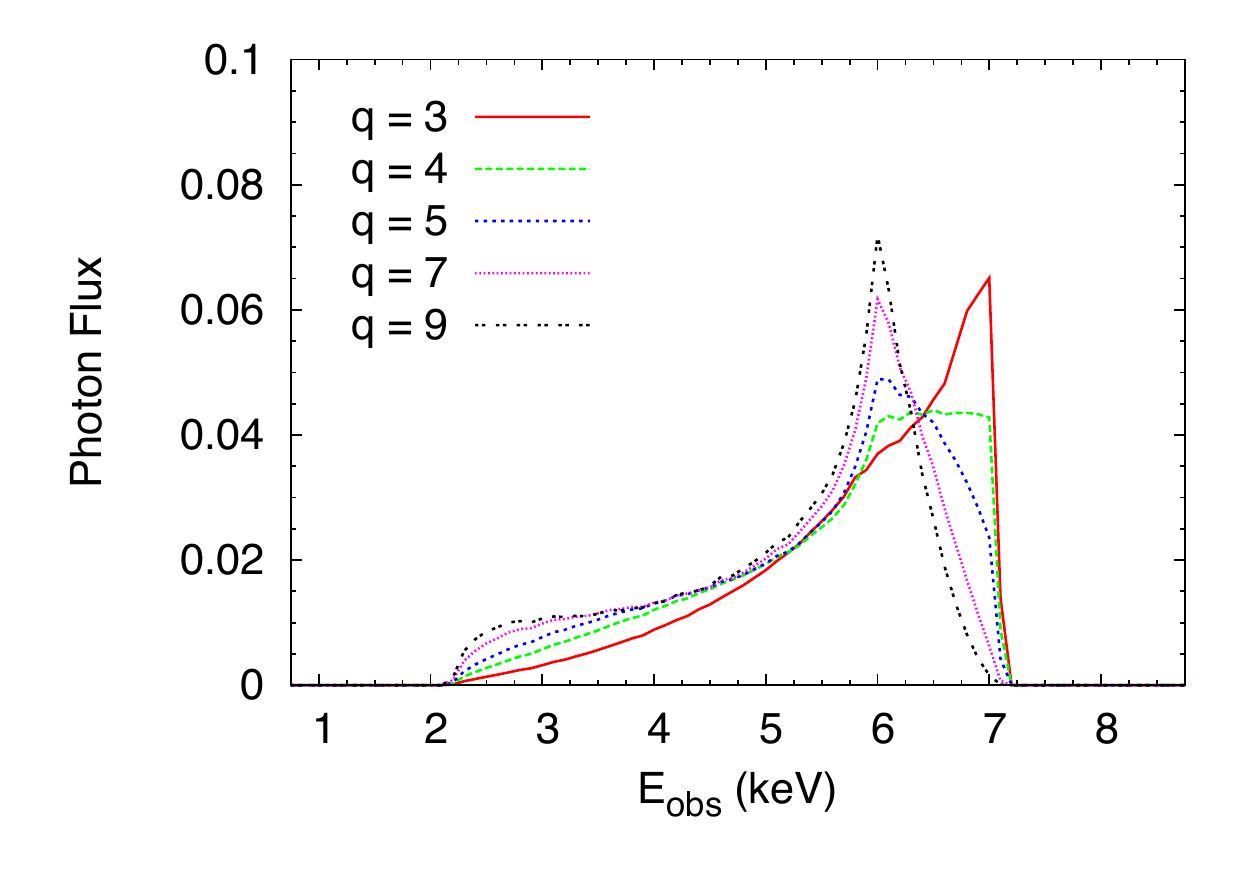} \\
\vspace{0.3cm}
\includegraphics[type=pdf,ext=.pdf,read=.pdf,width=8.5cm]{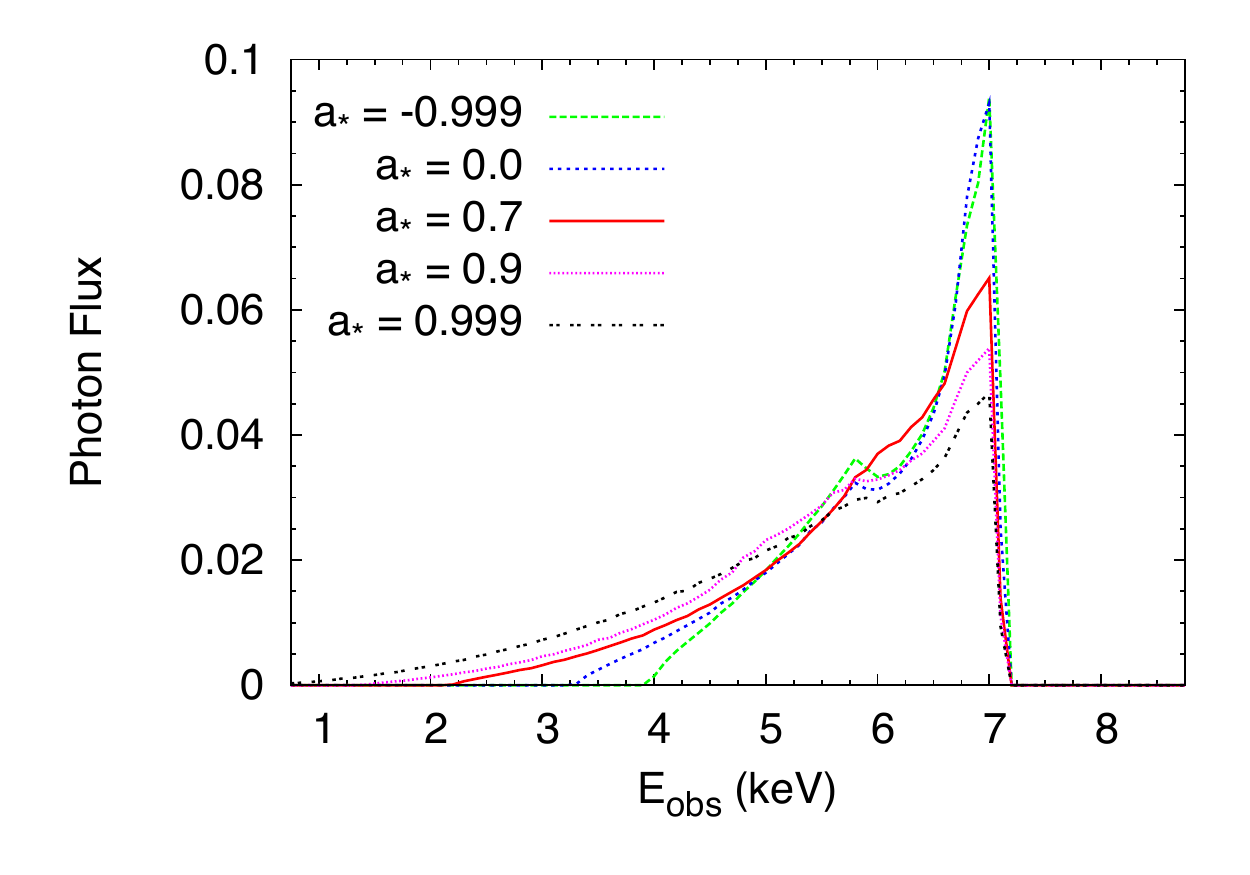}
\includegraphics[type=pdf,ext=.pdf,read=.pdf,width=8.5cm]{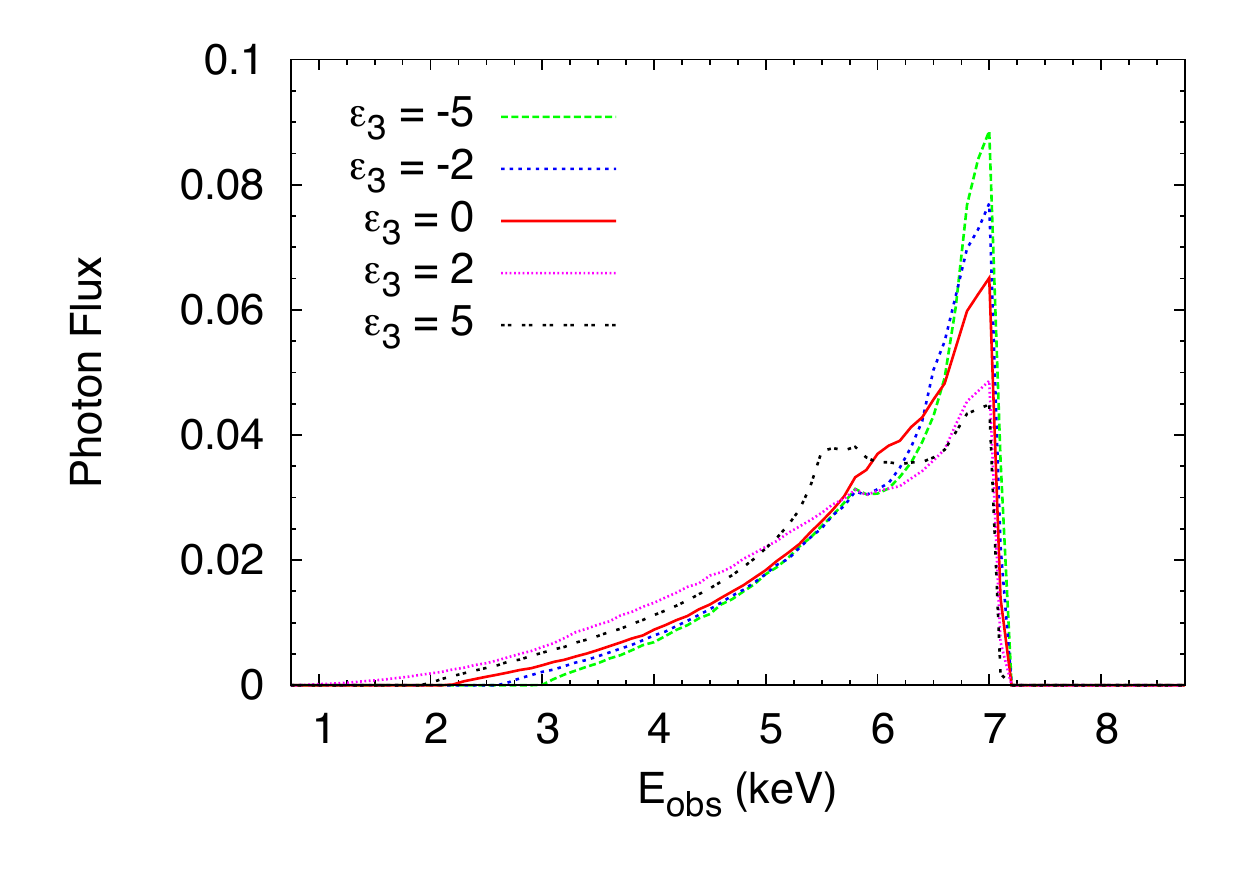}
\end{center}
\vspace{-0.5cm}
\caption{Impact of the model parameters on the iron line profile: viewing angle $i$ (top left panel), emissivity index $q$ (top right panel), spin parameter $a_*$ (bottom left panel), and Johannsen-Psaltis deformation parameter $\epsilon_3$ (bottom right panel). When not shown, the values of the parameters are: $i = 45^\circ$, $q = 3$, $a_*=0.7$, and $\epsilon_3 = 0$. The outer radius is $r_{\rm out} = r_{\rm ISCO} + 100$~$M$. Photon flux in arbitrary units.}
\label{f-iron1}
\end{figure*}

\subsection{Spin measurements with the iron line}

Since the shape of the iron line profile is the result of relativistic effects, its study can potentially be used to investigate the spacetime geometry around a black hole candidate. Assuming the Kerr metric, the technique can estimate the spin parameter $a_*$~\cite{fe1,fe2,brenneman-review13,fe3}. In practice, one models the full reflection spectrum although the iron K$\alpha$ line and edge are the most pronounced features.

The shape of the line is primarily determined by the background metric, the geometry of the emitting region, the disk emissivity, and the viewing angle of the disk. In the Kerr metric, the background has only one parameter, the spin $a_*$, while the mass of the black hole sets the scale of the system and does not affect directly the shape of the line. The emission region is the accretion disk, with the inner radius $r_{\rm in} = r_{\rm ISCO}$ and some outer radius $r_{\rm out}$. The latter is usually assumed large enough that its exact value does not matter. The intensity profile is a subtle point. It may be modeled as a power-law $I_{\rm e} \propto r^{-q}$. In the simplest case, the emissivity index $q$ is a constant to be determined by the fit. A slightly more sophisticated choice it to assume an intensity profile $I_{\rm e} \propto r^{-q}$ for $r < r_{\rm break}$ and $I_{\rm e} \propto r^{-3}$ for $r > r_{\rm break}$, where $q=3$ corresponds to the Newtonian limit in the lamppost geometry at large radii far from the X-ray source. With this choice, we have two free parameters for the emissivity profile, $q$ and $r_{\rm break}$, to be determined by the fit. Assuming that the hot corona is point-like and on the axis of the accretion disk, (or equivalently for any geometry which can be similarly specified) it is possible to compute the emissivity index $q = q(r)$ according to the height of the corona $h$~\citep{dauser}.

The impact of the viewing angle $i$, the emissivity index $q$, and the spin parameter $a_*$ on the iron line profile is shown in Fig.~\ref{f-iron1}. The viewing angle mainly affects the Doppler boosting, such that the high energy part of the profile shifts to higher energies as $i$ increases. The maximum blueshift is produced at 10-15 gravitational radii from the center, where the gas velocity is still high and the gravitational redshift is much weaker than at smaller radii. The emissivity index dictates the contribution from every radius on the disk. As $q$ increases, the radiation is more centrally concentrated.  Therefore, even the peak of the line moves to lower energies, since the radiation is more strongly affected by gravitational redshift. The main effect of the spin parameter is to change the position of the ISCO radius. If $a_*$ increases, a more significant fraction of the photons arriving at the detector are emitted very close to the black hole. Here the gravitational redshift is stronger and therefore the red tail of the iron line moves to lower energies.

A summary of current spin measurements of stellar-mass and supermassive black hole candidates with the iron line technique is shown in Tabs.~\ref{tab1} and \ref{tab2}. In the case of stellar-mass black hole candidates, some iron line measurements can be compared to the continuum-fitting results. They typically agree. For instance, in the case of Cygnus~X-1 and GRS~1915 both measurements find a high value of the spin parameter. For supermassive black hole candidates, the iron line is today much better suited for measuring the spin parameter of these objects and therefore it is not possible to cross-check the validity of the method. It is worth noting that supermassive black hole candidates are often found with high or extreme values of the spin parameter.

\begin{table*}[t]
\centering
\begin{tabular}{|ccccccc|}
\hline 
AGN & \hspace{0.1cm} & $a_*$ (iron) & \hspace{0.1cm} & $L_{\rm Bol}/L_{\rm Edd}$ & \hspace{0.1cm} & Principal References \\
\hline 
IRAS~13224-3809 && $> 0.995$ && $0.71$ && \cite{W13} \\
Mrk~110         && $> 0.99$  && $0.16 \pm 0.04$ && \cite{W13} \\
NGC~4051        && $> 0.99$  && $0.03$ && \cite{P11} \\
MCG-6-30-15     && $> 0.98$  && $0.40 \pm 0.13$ && \cite{BR06,Miniutti2007} \\
1H~0707-495     && $> 0.98$  && $\sim 1$ && \cite{Calle-Perez2010,W13,Zoghbi2010} \\
NGC~3783        && $> 0.98$  && $0.06 \pm 0.01$ && \cite{B11,P11} \\
RBS~1124        && $> 0.98$  && $0.15$ && \cite{W13} \\
NGC~1365        && $0.97^{+0.01}_{-0.04}$ && $0.06^{+0.06}_{-0.04}$ && \cite{Risaliti2009b,Risaliti2013,Brenneman2013} \\
Swift~J0501.9-3239 && $> 0.96$  && --- && \cite{W13} \\
Ark~564         && $0.96^{+0.01}_{-0.06}$ && $> 0.11$ && \cite{W13} \\
3C~120          && $> 0.95$  && $0.31 \pm 0.20$ && \cite{Lohfink2013} \\
Ark~120         && $0.94 \pm 0.01$ && $0.04 \pm 0.01$ && \cite{P12,Nardini2011,W13} \\
Ton~S180        && $0.91^{+0.02}_{-0.09}$ && $2.1^{+3.2}_{-1.6}$ && \cite{W13} \\
1H~0419-577     && $> 0.88$ && $1.3 \pm 0.4$ && \cite{W13} \\
Mrk~509         && $0.86^{+0.02}_{-0.01}$ && --- && \cite{W13} \\
IRAS~00521-7054 && $> 0.84$ && --- && \cite{Tan2012} \\
3C~382          && $0.75^{+0.07}_{-0.04}$ && --- && \cite{W13} \\
Mrk~335         && $0.70^{+0.12}_{-0.01}$ && $0.25 \pm 0.07$ && \cite{P12,W13} \\
Mrk~79          && $0.7 \pm 0.1$ && $0.05 \pm 0.01$ && \cite{Gallo2005,Gallo2011} \\
Mrk~359         && $0.7^{+0.3}_{-0.5}$ && $0.25$ && \cite{W13} \\
NGC~7469 && $0.69 \pm 0.09$ && --- && \cite{P12} \\
Swift~J2127.4+5654 && $0.6 \pm 0.2$ && $0.18 \pm 0.03$ && \cite{Miniutti2009a,P12} \\ 
Mrk~1018        && $0.6^{+0.4}_{-0.7}$ && $0.01$ && \cite{W13} \\
Mrk~841         && $> 0.56$  && $0.44$ && \cite{W13} \\
Fairall~9       && $0.52^{+0.19}_{-0.15}$ && $0.05 \pm 0.01$ && \cite{P12,L12,Schmoll2009,W13} \\
\hline 
\end{tabular}
\vspace{0.4cm}
\caption{Summary of the iron line measurements of the spin parameter of supermassive black hole candidates under the assumption of the Kerr background. See the references in the last column and~\cite{brenneman-review13} for more details. \label{tab2}}
\end{table*}

\begin{figure*}
\begin{center}
\vspace{0.5cm}
\includegraphics[type=pdf,ext=.pdf,read=.pdf,width=7.0cm]{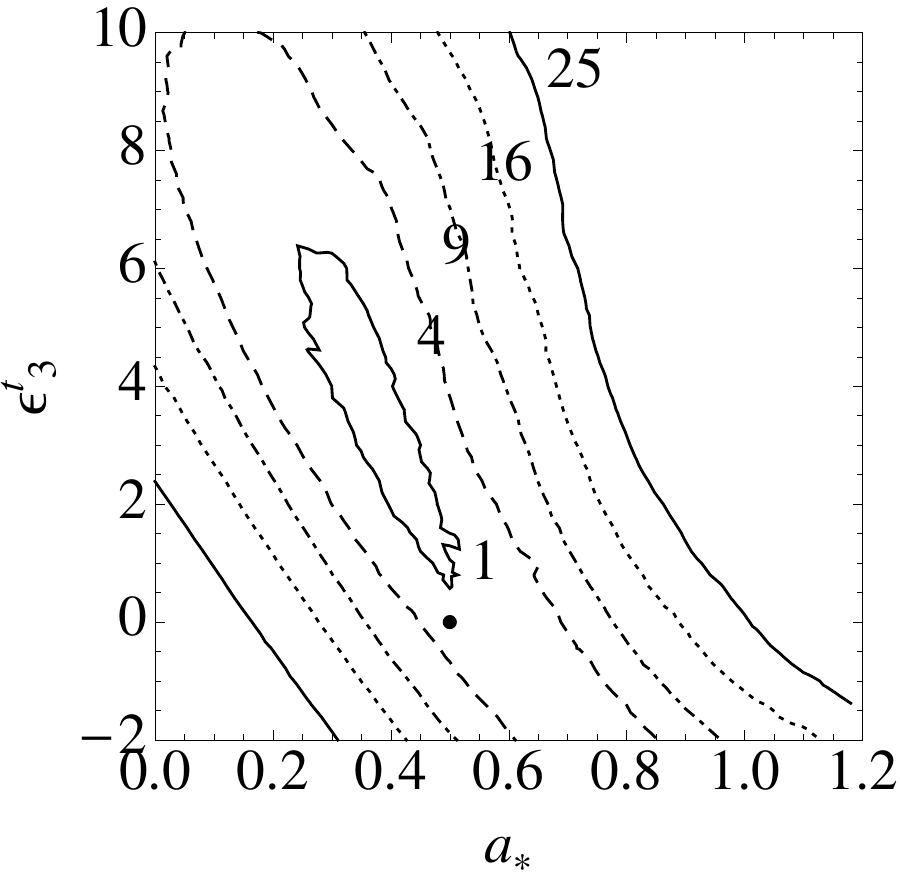}
\hspace{0.5cm}
\includegraphics[type=pdf,ext=.pdf,read=.pdf,width=7.0cm]{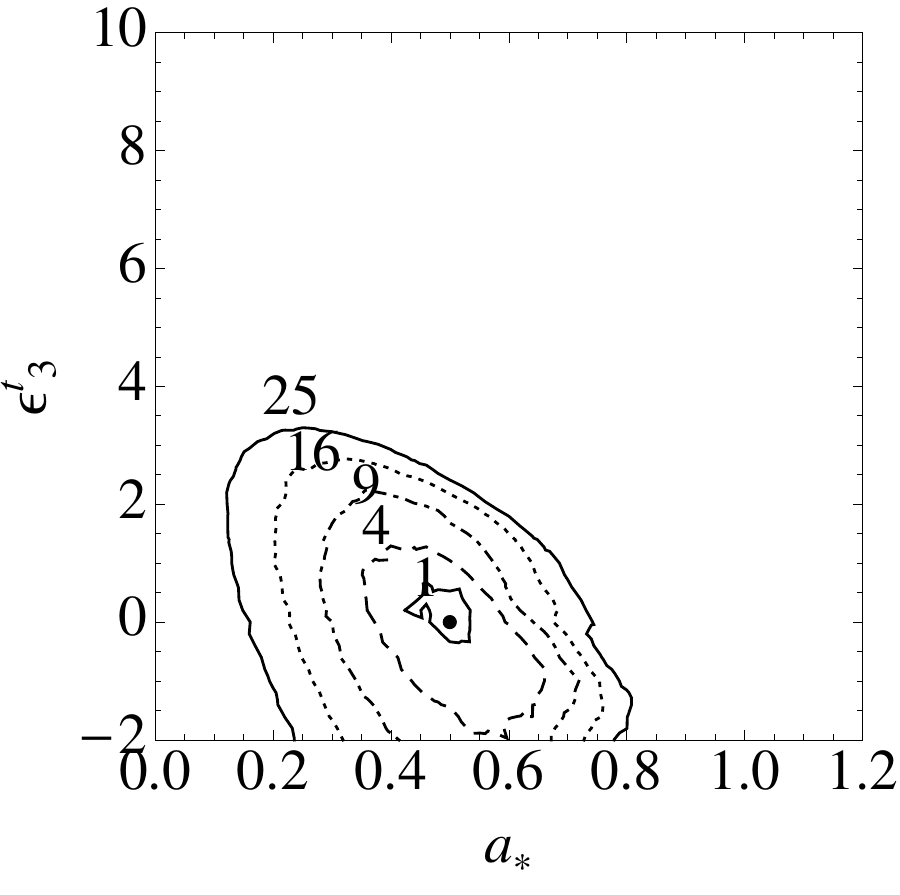} \\
\vspace{0.8cm}
\includegraphics[type=pdf,ext=.pdf,read=.pdf,width=7.0cm]{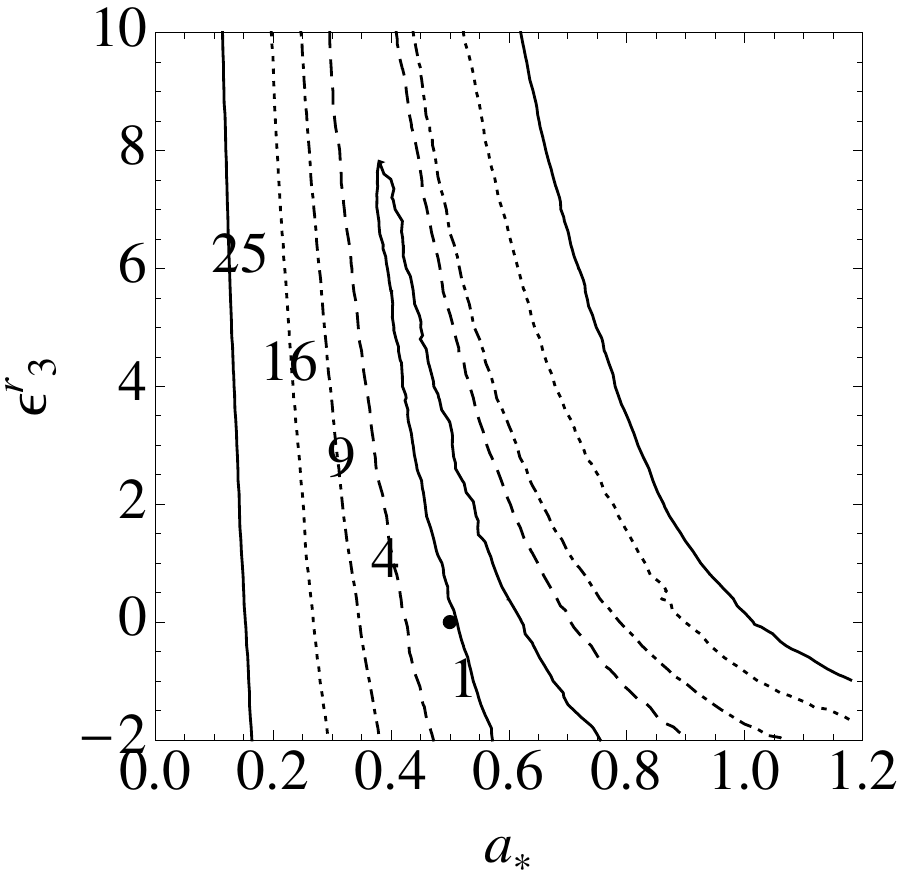}
\hspace{0.5cm}
\includegraphics[type=pdf,ext=.pdf,read=.pdf,width=7.0cm]{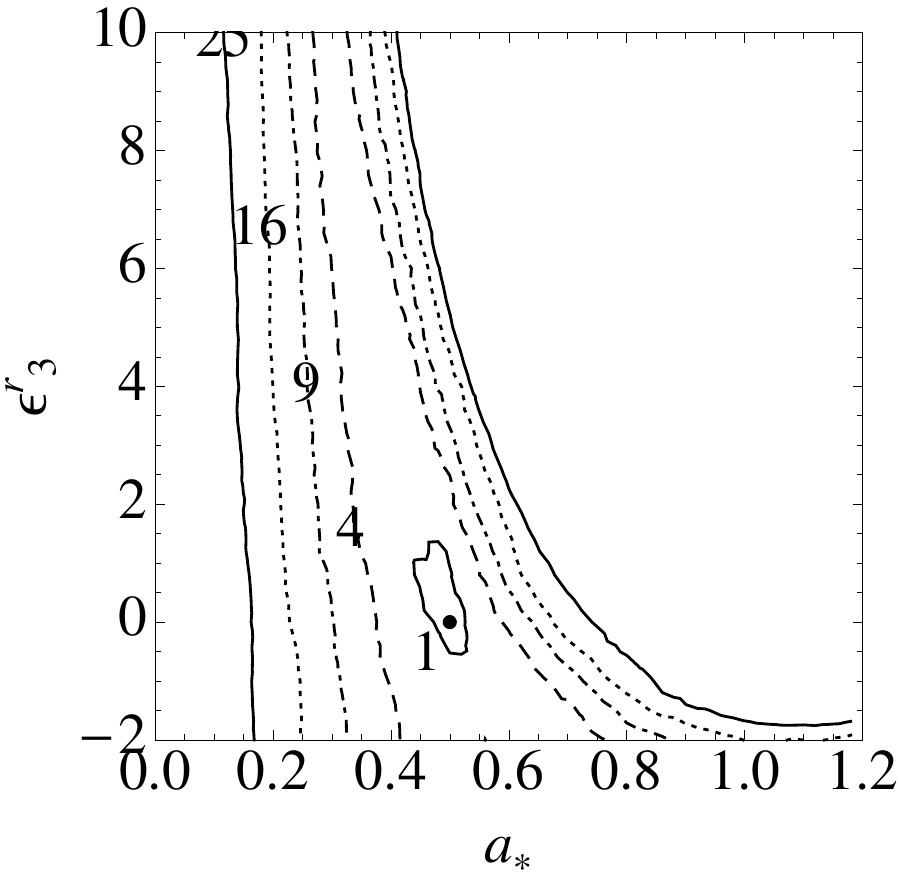}
\end{center}
\caption{Top panels: $\Delta \chi^2$ contours with $N=10^4$ from the comparison of the iron line profile of a Kerr black hole simulated using an input spin parameter $a_*' = 0.5$ and an inclination of $i'=20^\circ$ (left panel) and $i' = 70^\circ$ (right panel) versus a set of Cardoso-Pani-Rico black holes with spin parameters $a_*$ and a non-vanishing deformation parameter $\epsilon^t_3$. Bottom panels: as in the top panels for a non-vanishing deformation parameter $\epsilon^r_3$. For simplicity, $q$ is fixed to 3 and is not a fit parameter. From Ref.~\cite{jjc2}. }
\label{f-iron2}
\end{figure*}

\begin{figure*}
\begin{center}
\vspace{0.5cm}
\includegraphics[type=pdf,ext=.pdf,read=.pdf,width=7.0cm]{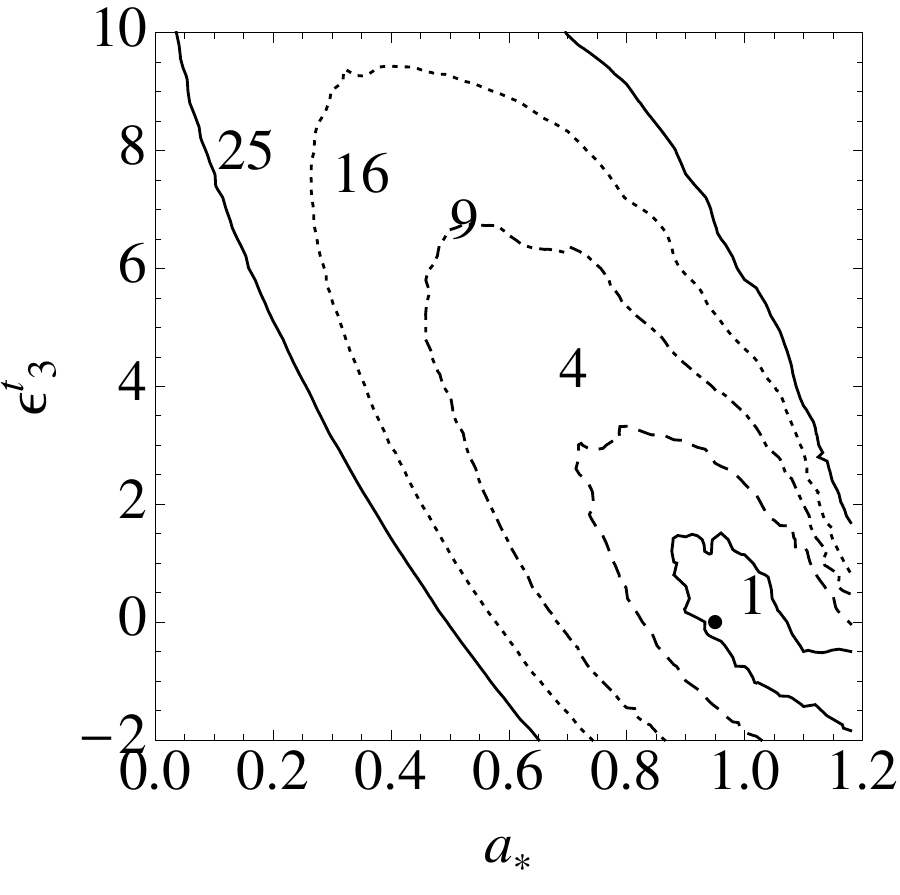}
\hspace{0.5cm}
\includegraphics[type=pdf,ext=.pdf,read=.pdf,width=7.0cm]{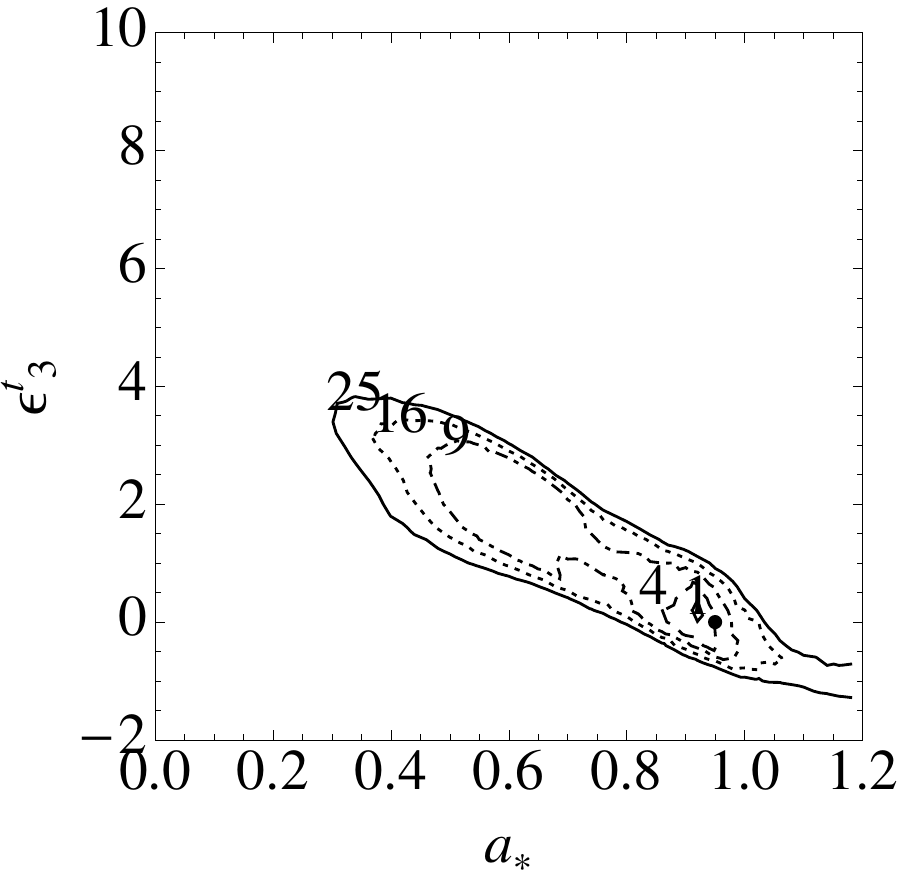} \\
\vspace{0.8cm}
\includegraphics[type=pdf,ext=.pdf,read=.pdf,width=7.0cm]{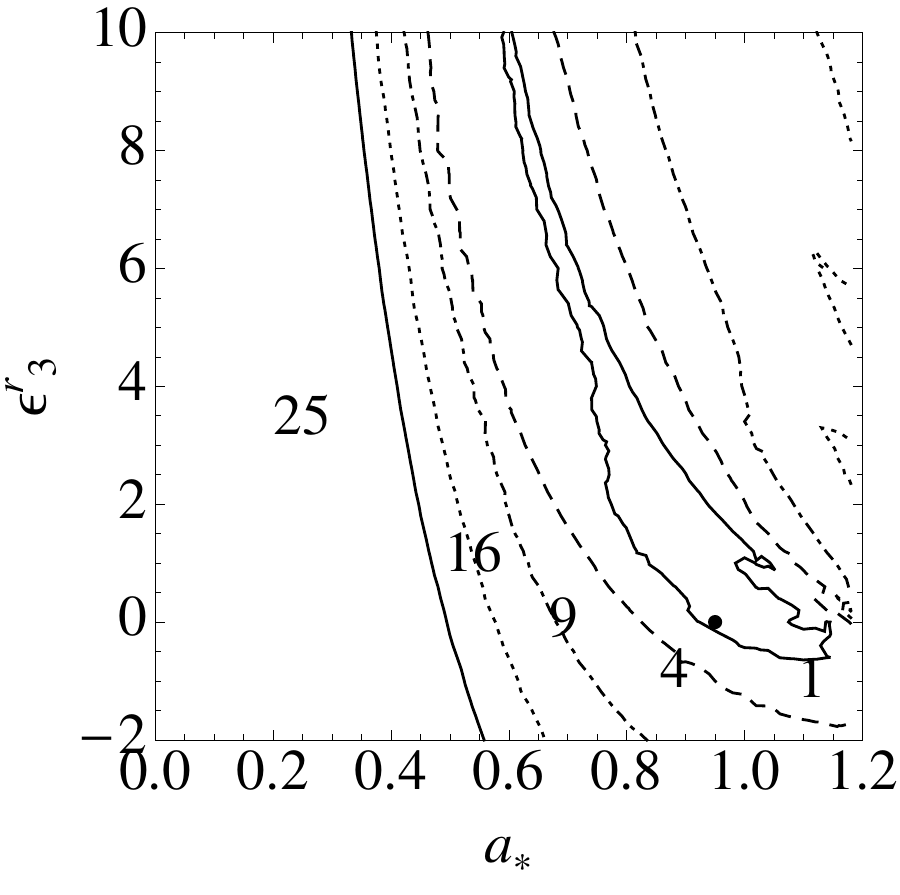}
\hspace{0.5cm}
\includegraphics[type=pdf,ext=.pdf,read=.pdf,width=7.0cm]{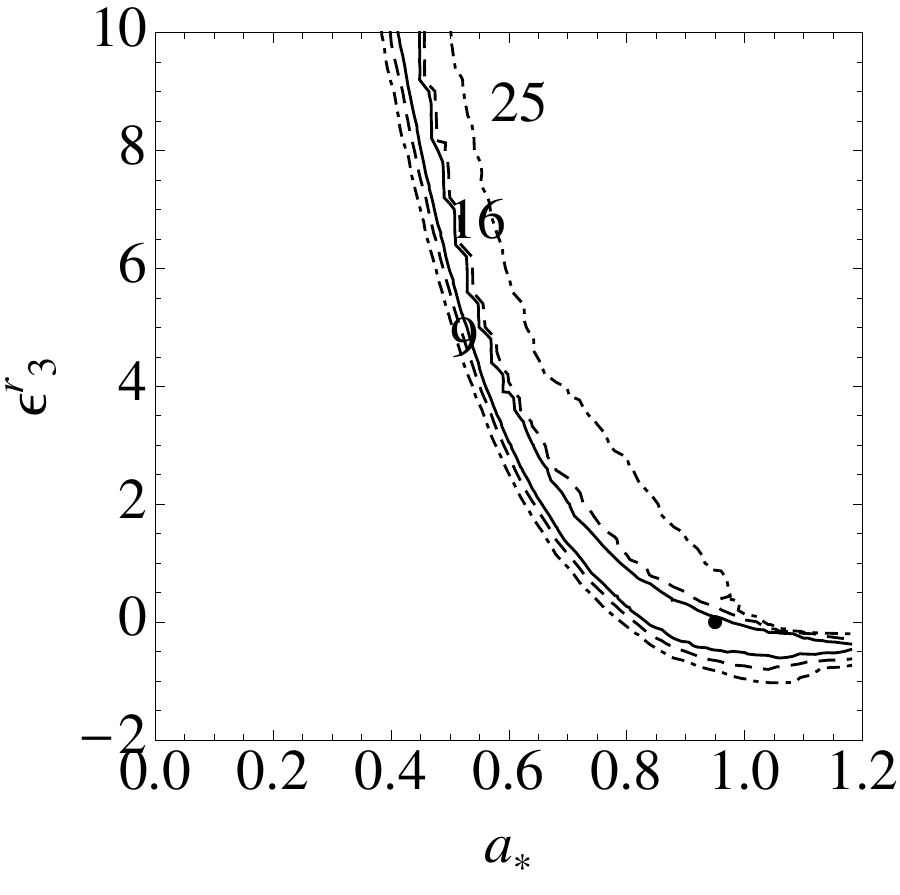}
\end{center}
\caption{Top panels: $\Delta \chi^2$ contours with $N=10^4$ from the comparison of the iron line profile of a Kerr black hole with an input spin parameter $a_*' = 0.95$ and an inclination of $i'=20^\circ$ (left panel) and $i' = 70^\circ$ (right panel) versus a set of Cardoso-Pani-Rico black holes with spin parameters $a_*$ and a non-vanishing deformation parameter $\epsilon^t_3$. Bottom panels: as in the top panels for a non-vanishing deformation parameter $\epsilon^r_3$. For simplicity, $q$ is fixed to 3 and is not a fit parameter. From Ref.~\cite{jjc2}. }
\label{f-iron3}
\end{figure*}

\begin{figure*}
\begin{center}
\vspace{0.5cm}
\includegraphics[type=pdf,ext=.pdf,read=.pdf,width=7.0cm]{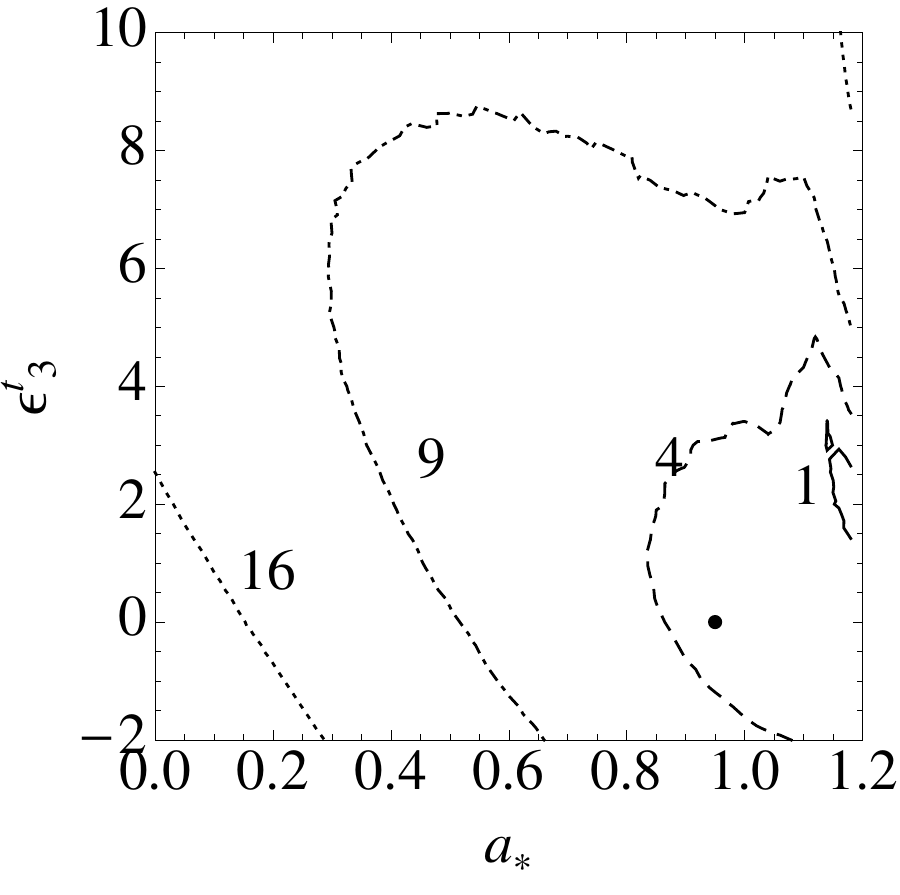}
\hspace{0.5cm}
\includegraphics[type=pdf,ext=.pdf,read=.pdf,width=7.0cm]{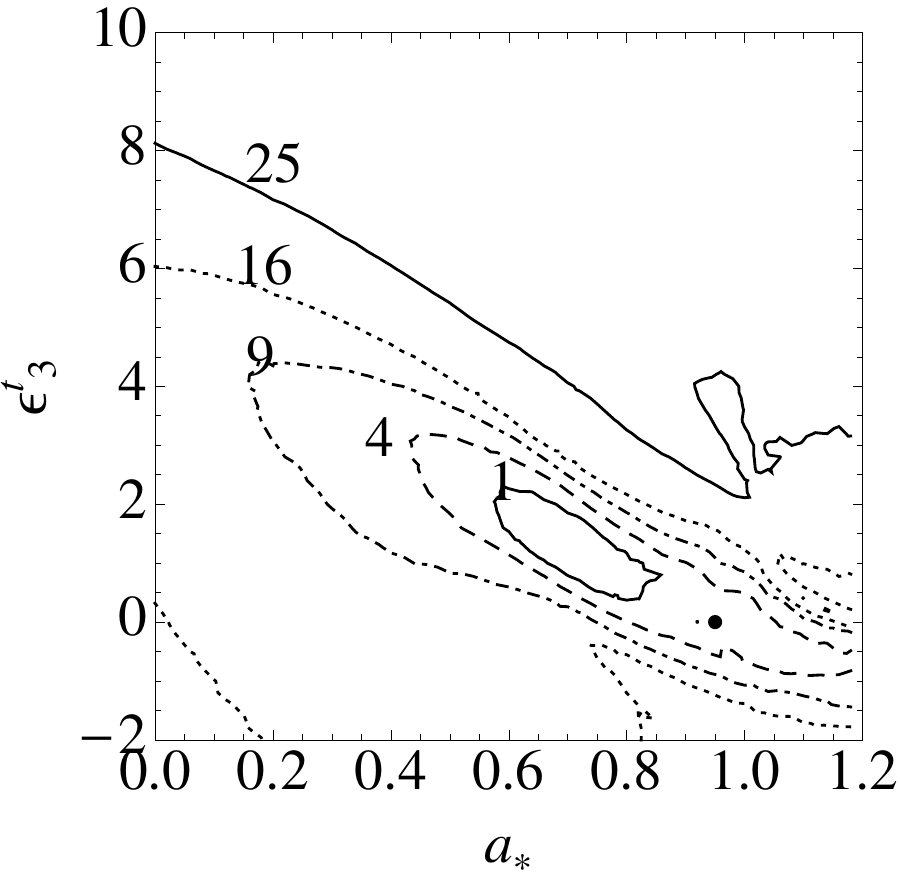} \\
\vspace{0.8cm}
\includegraphics[type=pdf,ext=.pdf,read=.pdf,width=7.0cm]{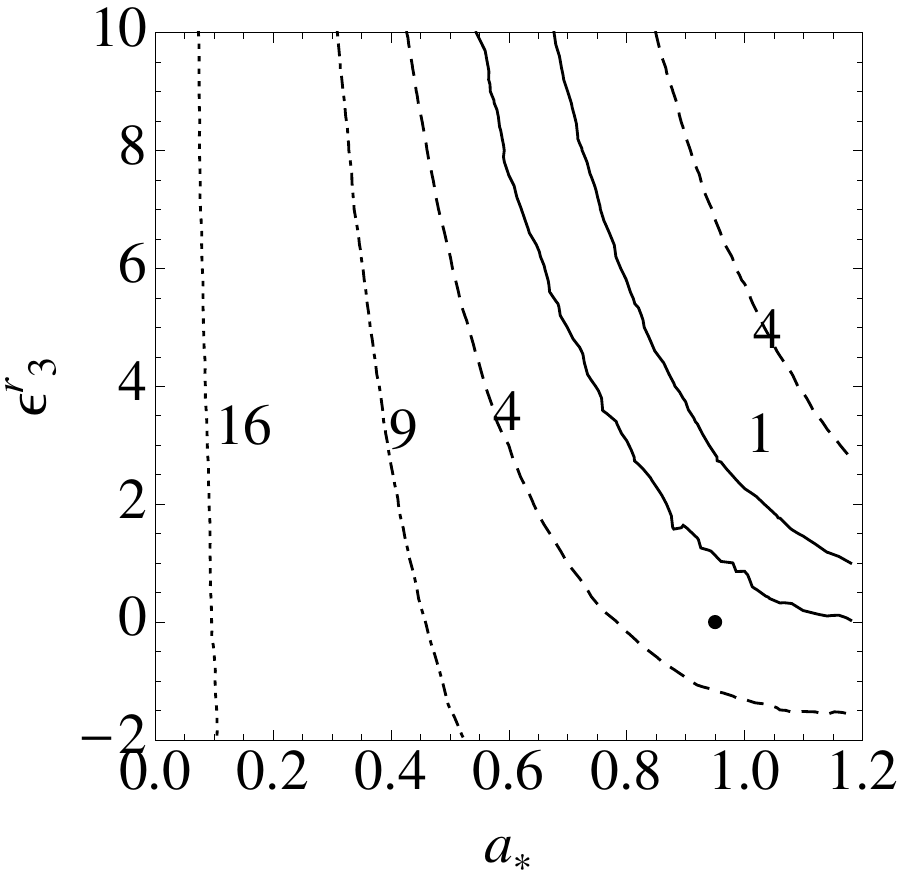}
\hspace{0.5cm}
\includegraphics[type=pdf,ext=.pdf,read=.pdf,width=7.0cm]{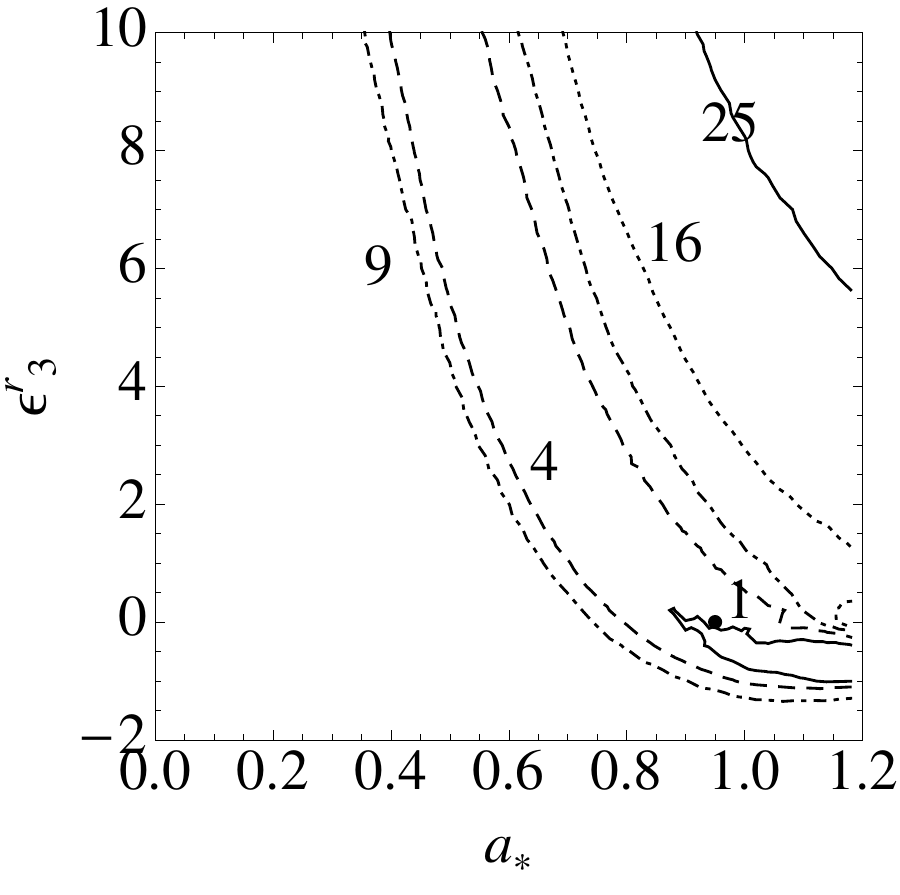}
\end{center}
\caption{As in Fig.~\ref{f-iron2} for $N=10^3$. From Ref.~\cite{jjc2}. }
\label{f-iron4}
\end{figure*}

\subsection{Testing black hole candidates with the iron line}

Since the shape of the profile of the iron line is determined by Doppler boosting, gravitational redshift, and light bending, the geometry of the spacetime affects the line shape, and its analysis  can potentially be used to test the Kerr metric~\cite{iron1,iron2,iron3,iron4,iron5,jjc1,jjc2}.

The impact of the Johannsen-Psaltis deformation parameter $\epsilon_3$ on the iron line profile is shown in the right bottom panel in Fig.~\ref{f-iron1}. Like the spin parameter $a_*$, the main effect of $\epsilon_3$ is to vary the ISCO radius. If $\epsilon_3$ becomes positive (negative), the radius of the ISCO decreases (increases) and thus the low energy tail of the line moves to lower (higher) energies with respect to the Kerr case with the same spin parameter. This is analogous to the continuum-fitting method. However, the iron line has a distinctive structure and its shape is not as simple as the that of the thermal spectrum of a thin disk.

The constraining power of the iron line has been explored in~\cite{jjc1,jjc2}. The conclusion of these papers is that, in the absence of high quality X-ray data (the circumstance of current observations), the continuum-fitting and the iron line methods provide similar constraints. In the case of the continuum-fitting method, the main source of uncertainty comes from measurements of the mass, distance, and inclination angle of the source, which typically come from optical observations. In the case of the iron line of AGN, the main limitation is the low photon count. In the presence of high quality X-ray data with a sufficiently high number of photon counts in the iron line, the iron line measurement can do a much better job and break the degeneracy between the spin and possible deviations from the Kerr solution (at least, provided we have a firm grasp of the corona's geometry).

Figs.~\ref{f-iron2}-\ref{f-iron4} show the constraints on $a_*$ and the Cardoso-Pani-Rico deformation parameters $\epsilon^t_3$ and $\epsilon^r_3$ obtained from simulations (see Ref.~\cite{jjc2} for the details). Fig.~\ref{f-iron2} shows the $\Delta\chi^2$ contours assuming an observation with with $N=10^4$ photons in the iron line. The simulation assumes the line contributes an equivalent width of $\sim 400$~eV, and includes a power-law continuum (with photon index $\Gamma=2$). It is for a Kerr black hole with spin parameter $a_*' = 0.5$ and inclination angle $i' = 20^\circ$ (left panels) and $70^\circ$ (right panels). In the top panels, $\epsilon^t_3$ may vary and $\epsilon^r_3=0$. The bottom panels show the opposite case, with $\epsilon^t_3 = 0$ and $\epsilon^r_3$ free. In all these figures, $q$ is fixed at 3.

Fig.~\ref{f-iron3} is like Fig.~\ref{f-iron2}, but now the simulations are for a Kerr black hole with spin parameter $a_*' = 0.95$. As in Fig.~\ref{f-iron2}, the deformation parameters can be better constrained when the inclination is high (which maximizes  relativistic effects) and $\epsilon^t_3$ is easier to constrain than $\epsilon^r_3$. A fast-rotating Kerr black hole is a more promising source to test the Kerr metric. This can be easily understood given the fact the inner edge of the disk is closer to the compact object and this, again, maximizes relativistic effects.

Fig.~\ref{f-iron4} is like Fig.~\ref{f-iron3}, but we assume a lower photon count in the line, $N=10^3$. This is roughly the signal available in present data of a bright AGN with current X-ray facilities. The constraints are definitively worse than those in Fig.~\ref{f-iron3}. At the moment, the limitation of the iron line measurement is set by the photon count and it is difficult to break the degeneracy between  spin and the deformation parameter. A sufficiently strong signal combined with an informed astrophysical model can achieve this test with next-generation instruments.

It is worth noting that, in the case of black hole binaries, there are additional complications to the reflection model. The degree of ionization in the disk is much higher, and so proper modeling of the ionization profile may be of greater importance (presently, this complication is generally ignored).  Further, the thermal component in the soft X-ray band produces additional Comptoniization in the disk and the thermal component itself can make modeling challenging for disentangling the contribution of the disk from the the low-energy tail of the iron line. ``Bare'' AGN may thus be more suitable for testing the Kerr metric with the iron line. However, Galactic black hole binaries provide the distinct advantage of being quite bright which is an important advantage given that observations of AGN are frequently limited in precision by the signal available in the broad iron line.

\subsection{Comparison between continuum-fitting and iron line methods}

In the case of stellar-mass black hole candidates, both the continuum-fitting and the iron line methods are applicable and, as shown in Tab.~\ref{tab1}, there are several objects that have been studied with both the techniques. Since parameter degeneracy between spin and possible deviations from the Kerr geometry is the main problem in tests of the Kerr metric, it is natural to examine whether the combinations of the two measurements for the same source can be helpful in breaking the parameter degeneracy~\cite{cfm-iron}.

As reviewed in Subsection~\ref{ss-cfm} and  is evident from Figs.~\ref{f-cfm2}-\ref{f-cfm4}, the continuum-fitting method measures the position of the ISCO radius and cannot obviously return more information because the shape of the spectrum is so simple that it is not possible to extract additional information. The iron line profile has a more complicated structure and it depends on several parameters that have to be fitted at the same time. This complexity is on the one hand cumbersome and on the other a crucial feature.  However, the main signature of the spacetime geometry is the extension of the low energy tail, which is directly related to the position of the ISCO radius: if the latter is closer to the compact object, the radiation emitted from the inner part of the accretion disk is more significantly affected by gravitational redshift and therefore the low energy tail extends to lower energies. In the end, in the presence of a typically weak signal, even the iron line achieves just a measurement of the ISCO radius and therefore the two techniques should obtain similar results and are not equipped to break the parameter degeneracy. Given sufficiently high-signal data, the iron line can potentially do a better job and break the  degeneracy.

The conclusion of the study in~\cite{cfm-iron} is thus that, in general with present data, the combination of the two approaches is not successful in proving departures from Kerr. If we assume the Kerr metric and we obtain the same spin measurement from both techniques, this is mere confirmation of the key assumption of these models, namely that the inner edge of the disk is at the ISCO radius.  Such confirmation may be directly useful for the methods, but it falls short of our more ambitious aim.  In the presence of high precision measurements, a discrepancy between the two values of the two estimates can arise (depending on the deviations from Kerr), but generally would require such precision and signal that the iron line data alone would provide its own constraint.  Given present systematic uncertainties, if the continuum-fitting and the iron line methods do not provide consistent measurements of the spin parameter (under the assumption of the Kerr background), then the first and most likely possibility to explore is in the astrophysical model.

\section{Summary and conclusions}

The continuum-fitting and the iron line methods are currently the principal techniques for probing the spacetime geometry around black hole candidates and testing the no-hair theorem. Other approaches, like measurements of quasi-periodic oscillations~\cite{qpo1,qpo2,qpo3,v2-mgpsf15}, observation of a pulsar orbiting a black hole candidate~\cite{pul1,pul2}, or detection of the black hole shadow~\cite{sha1,sha2,sha3}, may be possible in the future.

The continuum-fitting and the iron line methods were originally proposed and developed to measure the spin parameter of black hole candidates under the assumption that their spacetime metric is described by the Kerr solution. The generalization to test black hole candidates is quite straightforward. This approach, strictly speaking, does not test general relativity, because the properties of the electromagnetic radiation emitted by the gas in the disk depends on the motion of the particle of the gas in the disk and  the propagation of the radiation from the disk to the detector -- namely it only depends on the background metric -- and it cannot distinguish a Kerr black hole in general relativity from a Kerr black hole in an alternative theory of gravity, as the geodesic motion is the same. The approach in question tests the validity of the Kerr metric.

The continuum-fitting method consists of analysis of the thermal spectrum of a geometrically thin and optically thick accretion disk. The technique is relatively robust and well understood. In the Kerr metric, there is a one-to-one correspondence between the spin and the disk radiative efficiency/ISCO radius, and this permits us to estimate the spin parameter. If we relax the Kerr black hole hypothesis, there is typically a degeneracy between the spin and possible deviations from the Kerr solution, with the result that, in general, the continuum-fitting method cannot, by itself, test the Kerr metric. However, it is possible for this approach to constrain certain deviations, because very deformed objects may have a low radiative efficiency and therefore the observation of a black hole candidate with high efficiency (i.e., like a fast-rotating Kerr black hole) may rule out such deformations.

The analysis of the iron line profile is potentially a more powerful probe, but the astrophysical model is more complicated and a rigorous model requires deeper knowledge of the corona, the nature of which is poorly understood at present.  Furthermore, for AGN, an order of magnitude (or more) increase in the signal of typical spectral data is needed.  The main impact of the spacetime geometry on the iron line is on the redward extent of the low energy tail.  This, in turn, is linked to the ISCO radius, as well as other ``hair'' which may be present. With current X-ray data, the continuum-fitting and the iron line measurements seem to be able to provide comparable constraints.  With better quality data in the near future, and with a better handle on the systematics, great advances may be possible, particularly using the rich information imprinted by reflection in the strong-gravity domain.


\begin{acknowledgments}
C.B. and J.J. were supported by the NSFC grants No.~11305038 and No.~U1531117, the Shanghai Municipal Education Commission grant No.~14ZZ001, the Thousand Young Talents Program, and Fudan University. C.B. acknowledges also support from the Alexander von Humboldt Foundation. J.F.S. was supported by the NASA Einstein Fellowship grant PF5-160144.
\end{acknowledgments}


\end{document}